\documentclass[preprint,12pt]{elsarticle}
\usepackage[version=4]{mhchem}
\usepackage[T1]{fontenc}
\usepackage[utf8]{inputenc}
\usepackage{geometry}
\usepackage{verbatim}
\usepackage{amsmath}
\usepackage{amssymb}
\usepackage{graphicx}
\usepackage{setspace}
\usepackage{physics}
\usepackage{chngcntr}
\usepackage{etoolbox}
\usepackage{algpseudocode}
\usepackage{caption}
\usepackage{mathtools}
\usepackage{algorithm}
\usepackage[dvipsnames]{xcolor}
\usepackage[normalem]{ulem}
\usepackage{cancel}
\makeatletter
\providecommand{\tabularnewline}{\\}
\usepackage{float}
\usepackage{makecell}
\date{}

\usepackage[english]{babel} 

\@ifundefined{showcaptionsetup}{}{%
 \PassOptionsToPackage{caption=false}{subfig}}
\usepackage{subfig}
\makeatother

\usepackage{babel}

\usepackage[hidelinks]{hyperref}
\hypersetup{
    colorlinks=false,
}
\makeatletter
\def\ps@pprintTitle{%
  \let\@oddhead\@empty
  \let\@evenhead\@empty
  \let\@oddfoot\@empty
  \let\@evenfoot\@oddfoot
}
\makeatother
\journal{XYZ}

\usepackage{enumitem}
\usepackage{xurl}            
\newenvironment{programsummary}{%
  \setlength{\parindent}{0pt}
  \setlength{\parskip}{0.5ex}
  \small
}{%
  \smallskip\noindent\hrule\par\smallskip
}

\begin{document}
\begin{frontmatter}
\title{\texttt{CuPyMag}: GPU-Accelerated Finite-Element Micromagnetics with Magnetostriction}

\author[inst1]{Hongyi Guan\corref{cor1}}
\cortext[cor1]{The first author maintains [url] and is the primary contact for technical or code-related issues.}
\ead{hongyi_guan@ucsb.edu}

\author[inst1]{Ananya Renuka Balakrishna\corref{cor2}}
\ead{ananyarb@ucsb.edu}
\cortext[cor2]{Corresponding author}
\affiliation[inst1]{organization={Materials Department},
            addressline={University of California, Santa Barbara}, 
            city={Santa Barbara},
            postcode={93106}, 
            state={CA},
            country={USA}}

\begin{abstract}
We introduce \texttt{CuPyMag}, an open-source, Python-based framework for large-scale micromagnetic simulations with magnetostriction. \texttt{CuPyMag} solves micromagnetics with finite elements in a GPU-resident workflow in which key operations, such as right-hand-side assembly, spatial derivatives, and volume averages, are tensorized using CuPy’s BLAS-accelerated backend.
Benchmark tests show that the GPU solvers in \texttt{CuPyMag} achieve a speedup of up to two orders of magnitude compared to the CPU codes. Its runtime grows linearly/sublinearly with problem size, demonstrating high efficiency. Additionally, \texttt{CuPyMag} uses the Gauss-Seidel projection method for time integration, which not only allows stable time steps (up to 11 ps) but also solves each governing equation with only 1–3 conjugate-gradient iterations without preconditioning. \texttt{CuPyMag} accounts for magnetoelastic coupling and far-field effects arising from the boundary of the magnetic body, both of which play an important role in magnetization reversal in the presence of local defects. \texttt{CuPyMag} solves these computationally-intensive multiphysics simulations with a high-resolution mesh (up to 3M nodes) in under three hours on an NVIDIA H200 GPU. 
This acceleration enables micromagnetic simulations with non-trivial defect geometries and resolves nanoscale magnetic structures. It expands the scope of micromagnetic simulations towards realistic, large-scale problems that can guide experiments. More broadly, \texttt{CuPyMag} is developed using widely adopted Python libraries, which provide cross-platform compatibility, ease of installation, and accessibility for adaptations to diverse applications.
\end{abstract}


\end{frontmatter}
\section*{Program summary}
\begin{programsummary}
\textit{Program Title:} \texttt{CuPyMag}

\textit{CPC Library link to program files:} [url]

\textit{Developer’s repository link:} [url]

\textit{OSF repository link:} [url]

\textit{Licensing provisions:} Apache-2.0

\textit{Programming language:} Python 3

\textit{Nature of problem:} The program addresses large-scale micromagnetic simulations that are computationally demanding due to the solution of magnetoelastic coupling (mechanical equilibrium with magnetostriction at each time step), far-field demagnetization effects, and arbitrarily shaped inclusion-like defects that require unstructured finite-element meshes.

\textit{Solution method:} \texttt{CuPyMag} implements a GPU-resident, tensorized workflow built on CuPy’s BLAS-accelerated backend. This design exploits highly optimized linear algebra kernels, enabling efficient parallel execution and maintaining high GPU utilization. It uses an ellipsoid theorem to account for far-field demagnetization effects, and the Gauss-Seidel projection method for stable and efficient time integration.

\end{programsummary}

\section{Introduction}
Micromagnetic simulations are widely used to study magnetization dynamics and domain structures at an intermediate scale between individual atoms and the macroscale \cite{brown1963micromagnetics}. Unlike density functional theory, which provides quantum mechanical insight at the atomic scale, or Maxwell’s equations, which govern large-scale electromagnetic behavior, micromagnetics focuses on a mesoscopic regime. At this length scale, micromagnetics theory predicts the intricate domain patterns and dynamics that arise from the interplay between magnetic dipoles, crystal anisotropy, and external magnetic fields. The nucleation and growth of these domain patterns affect the material response, e.g., hysteresis, actuation, and are critical to the understanding and optimizing of magnetic behavior in various applications \cite{PhysRevMaterials.9.044407,Okamoto2021,Nakai2024}. These properties are essential for the design of spintronic devices, high-density magnetic storage, magnetic sensors \cite{Fidler2000,hubert1998magnetic,10.1063/1.5093730}, and emerging technologies such as neuromorphic computing \cite{molecules25112550} and nanomagnetic logic \cite{Carlton2012Nanomagnetic}. 

Micromagnetics involves solving the minimization of the total free energy of a magnetized body with respect to a magnetization vector field, $\vb{M}$ \cite{brown1963micromagnetics, brown1966magnetoelastic}. This minimization involves numerically solving partial differential equations (PDEs) for magnetostatic and mechanical equilibrium conditions, as well as magnetization dynamics described by the Landau-Lifshitz-Gilbert (LLG) equation. Analytical solutions for these PDEs are, however, often confined to simplified linearized models \cite{10.1063/1.2185970, Leliaert_2018}, and numerical calculations greatly generalize and diversify the magnetism problems to study, such as investigating the topology of a magnetic skyrmion on a thin film \cite{GERGIDIS2019111} and exploring magnetization dynamics during spin wave computing \cite{10.1063/5.0019328}. These calculations can be expensive due to several factors, including the need to resolve fine magnetic structures (e.g., domain walls, vortices) \cite{Andreas2014StrongInhomogeneities}, time integration of the non-linear LLG equations \cite{Suess2002PreconditionedLLG}, computing the long-range demagnetization field \cite{Lebecki2008PBCDemag}, and accounting for the multiphysics coupling in the system (e.g., magnetoelastic interactions) \cite{Reichel2023EfficientMicromagneticElastic}. By addressing these computational challenges using high-performance computing, we accelerate materials exploration \cite{chen2024accelerating}. Computationally efficient numerical micromagnetics, together with experimental efforts, can enable materials exploration across a wider parameter space, and thereby accelerate materials discovery.

A rich ecosystem of open-source packages has been developed for micromagnetic simulations (see Table~\ref{Tab:Micromagnetics_compare} and Ref.~\cite{10.1063/1.5093730}). Among these packages, finite-difference method (FDM) based tools such as \texttt{OOMMF} \cite{OOMMF} and \texttt{MuMax3} \cite{mumax3} are widely used due to their user-friendliness and high performance. \texttt{OOMMF} (NIST’s Object Oriented MicroMagnetic Framework) solves micromagnetics on a Cartesian grid with fast Fourier transform (FFT) method, but is limited to CPU execution. In contrast, several modern packages, such as \texttt{MuMax3} \cite{mumax3} and \texttt{magnum.np} \cite{Bruckner2023} implement GPU-accelerated FDM. Operations such as FFT and sparse matrix-vector multiplications (SpMV) on GPUs can significantly outperform their CPU counterparts of comparable generation \cite{5395212,electronics9101675}, and are promising for large-scale micromagnetic simulations \cite{Leliaert_2018,mumax3,Jermain2016}. Notably, \texttt{MuMax3} was benchmarked to scale up to 100M FFT cells, which is large enough to approach realistic device scales \cite{10.1063/1.5093730}.

However, FDM is inherently limited in modeling arbitrary shapes and boundary conditions, as non-orthogonal domains become ``staircased'', which in turn leads to spurious surface effects such as artificial ``magnetic charges'' along staircased boundaries \cite{NmagFAQ}. In physical systems, such boundaries often appear as defect interfaces, polycrystalline textures, and external sample boundaries. These shortcomings can be addressed by solving micromagnetics with the finite-element method (FEM), which handles complex geometries and boundary conditions more naturally. Additionally, FEM allows local mesh refinements, which contribute to increased resolution in regions with rapidly varying fields without substantially increasing the overall computational cost. 

Open-source FEM packages such as \texttt{Nmag} \cite{Fischbacher2007} and \texttt{magnum.fe} \cite{ABERT201329} have demonstrated the flexibility of unstructured meshes for curved and irregular shapes. However, these packages are built on CPUs and can only handle a limited range of problems due to the constrained parallelism and memory bandwidth of CPU architectures. To the best of our knowledge, open-source micromagnetic simulation codes are either GPU-accelerated but under the geometric constraints of FDM or allow flexible geometries through FEM but remain CPU-bound. As a result, users must choose between the high-throughput GPU FDM solvers and the geometrically versatile CPU-bound FEM codes, without a GPU-accelerated FEM framework to bridge the gap.

\begin{table}[t]
    \centering
    \begin{tabular}{c|cccc}
        Package & Discretization & Use GPU & \makecell{Programing \\ language}  \\ \hline
        \texttt{OOMMF} \cite{OOMMF} & FDM & No & C++  \\
        \texttt{MuMax3} \cite{mumax3} & FDM & Yes & Go  \\
        \texttt{magnum.np} \cite{Bruckner2023} & FDM & Yes & Python   \\
        \texttt{magnum.fd} \cite{magnumfd} & FDM & Yes & C++, Python   \\
        \texttt{NeuralMag} \cite{abert2024neuralmagopensourcenodalfinitedifference} & FDM & Yes & Python   \\
        \texttt{Nmag} \cite{Fischbacher2007} & FEM & No  & OCaml   \\
        \texttt{magnum.fe} \cite{ABERT201329} & FEM & No  & Python, C++  \\
        \texttt{Finmag} \cite{bisotti2018finmag} & FEM & No  & Python, C++  \\
        \texttt{TetraX} \cite{TetraX,korberFiniteelementDynamicmatrixApproach2021a} & FEM & No & Python, C \\
        \texttt{FeeLLGood} \cite{feellgood,Alouges2014} & FEM & No & C++, Python\\
        \texttt{CuPyMag} [This work] & FEM & Yes & Python
    \end{tabular}
    \caption{A summary of select open-source micromagnetic simulation packages available today. These packages use discretization methods that are either FDM (finite difference method) or FEM (finite element method). To our knowledge, no existing open-source micromagnetic simulation package incorporates a comprehensive magnetoelastic coupling, highlighting a critical limitation that is addressed in this work.}
    \label{Tab:Micromagnetics_compare}
\end{table}

We address this limitation by developing \texttt{CuPyMag}, an open-source GPU-accelerated FEM micromagnetic simulation framework. The key highlight of \texttt{CuPyMag} is its GPU-resident design, which requires only a one-time CPU-to-GPU data transfer for setup, while all subsequent simulation steps run entirely on the GPU. A CPU routine using Numba just-in-time (JIT) compilation \cite{10.1145/2833157.2833162} performs a one-time matrix assembly. Subsequently, the governing micromagnetics equations are solved entirely on the GPU through CuPy’s Basic Linear Algebra Subprograms (BLAS)-accelerated tensor operations \cite{cupy_learningsys2017}, which parallelizes across thousands of GPU cores. This design minimizes CPU-GPU (host-device) communication overhead. Moreover, \texttt{CuPyMag} uses the Gauss-Seidel projection method (GSPM) \cite{wang2001gauss} to solve the LLG equation, ensuring stable time integration while only requiring solving several well-conditioned Poisson-type equations. As a result of these design choices, \texttt{CuPyMag} completed a hysteresis calculation with 3M nodes in under 3 hours on a single H200 GPU, achieving a 2–3$\times$ speedup in double precision compared to an A100 GPU. This result shows that \texttt{CuPyMag}, despite being implemented entirely in Python, delivers high performance for large-scale materials simulations.

Besides its computational efficiency, \texttt{CuPyMag} offers several advantages in its theoretical formulation. First, \texttt{CuPyMag} implements the micromagnetics energy with magnetostriction, i.e., it accounts for coupled magnetoelastic interactions in which magnetization and strain fields mutually affect each other. In modeling this interaction, \texttt{CuPyMag} solves for mechanical equilibrium at each time step and ensures that the evolution of the strain field is consistent with that of the magnetization field. Leveraging its computational efficiency, \texttt{CuPyMag} can carry out this otherwise expensive mechanical equilibrium calculation on dense FEM meshes. These fully magnetoelastic coupled treatments are not common in the other open-source codes listed in Table.~\ref{Tab:Micromagnetics_compare}.

Recent studies have shown that neglecting this magnetoelastic coupling effect may lead to inaccuracies when analyzing complex magnetic domains with nonuniform magnetization fields in functional materials such as magnetostrictive alloys and magnetic shape memory alloys \cite{microstructures.2024.106}. Moreover, magnetoelastic coupling enables surface acoustic waves to excite magnetization dynamics and subsequent spin-wave propagation through adjacent waveguides \cite{10.1063/5.0260092}. This effect is also central to applications in spintronic sensors \cite{ma12071135,https://doi.org/10.1002/pssa.201000738,Ito2025}, actuators \cite{act8020045}, magnetoelectric laminates \cite{https://doi.org/10.1002/pssa.201000738}, and soft matter systems \cite{doi:10.1126/sciadv.ads0071}. Our previous work showed the subtle but important role of magnetoelasticity on hysteresis and provided new insights into the permalloy problem \cite{ARB2021Permalloy,balakrishna2021tool}. By carefully designing magnetoelastic interactions, we showed that soft magnets with large magnetocrystalline anisotropy can have small hysteresis \cite{renuka2022design,PhysRevMaterials.9.044407}.

In addition to magnetoelastic interactions, \texttt{CuPyMag} models inclusion-like defects with curved geometries. These nanoscale defects act as local perturbations to an otherwise uniform magnetization, serving as domain nucleation sites and pinning domain-wall motion, both of which alter the shape and size of hysteresis loops. Recent studies show that carefully engineered defect geometries can markedly reduce hysteresis losses in soft magnets, which is critical for energy efficiency \cite{han2022mechanically,han2023strong}. More broadly, such defect-induced domain wall motions and pinnings are important in permanent magnets used for motors and wind turbines \cite{Li2021CoercivityFAQ,Fischbacher_2018}, and spintronic devices such as magnetic RAMs and racetrack memory \cite{doi:10.1126/science.1145799}. In \texttt{CuPyMag}, users can design orthogonal (and non-orthogonal) defect geometries of inclusion-like or pore-like defects embedded inside the magnetic domain, apply finite boundary conditions on defect surfaces, and study their effect on macroscopic magnetic response. These examples highlight the broad significance of defect modeling in \texttt{CuPyMag}.

Finally, \texttt{CuPyMag} employs the ellipsoid theorem, a multiscale modeling strategy to account for local and long-range field interactions. In doing so, it predicts the effect of the demagnetization field originating from not only local imperfections or defects within the magnetic body, but also from the overall geometry of the magnetic body, on magnetization reversal. These features distinguish \texttt{CuPyMag} from other open-source micromagnetics solvers, providing users with a comprehensive computational tool for exploring magnetic behavior.

In the following sections, we outline the micromagnetics theory and the governing equations implemented in \texttt{CuPyMag} and the numerical implementation within the GPU-resident workflow. We then benchmark the key components of the workflow, including matrix assembly, sparse linear solver, and LLG time-integration efficiency, as well as the overall runtime. We analyze performance trends across different problem sizes, single precision vs. double precision, and the GPU hardware (A100 vs. H200). Building on these results, we demonstrate \texttt{CuPyMag}'s capability to efficiently solve representative micromagnetic simulations of predicting hysteresis loops and domain patterns in soft magnetic alloys with defects. Finally, we discuss the significance and limitations of \texttt{CuPyMag} and outline potential directions for future code development.

\section{Theory}
\begin{figure}[!t]
    \centering
    \includegraphics[width=\linewidth]{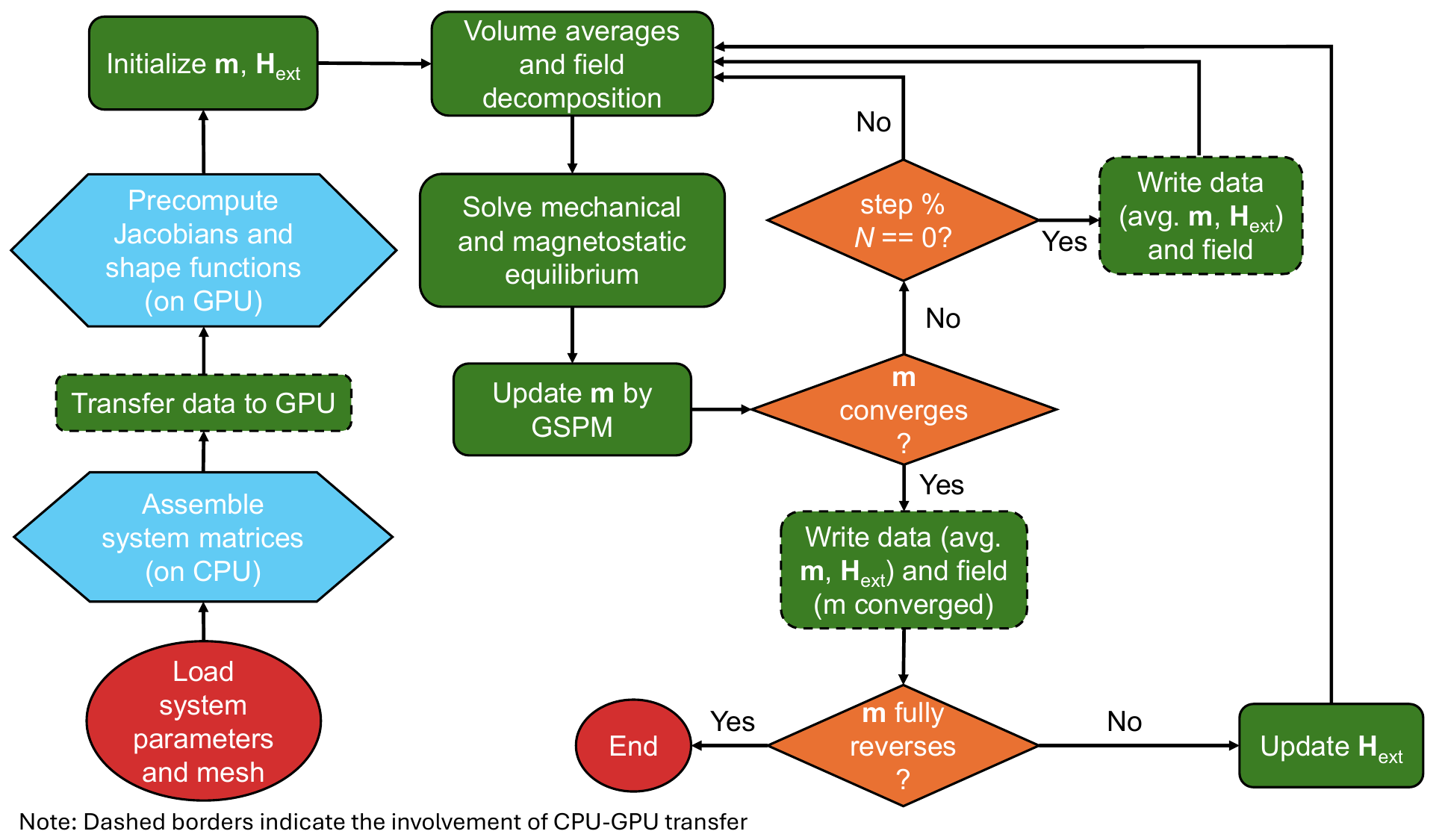}
    \caption{The complete workflow of \texttt{CuPyMag}. Here, the average magnetization, the external fields $\mathbf{H}_{\mathrm{ext}}$, and field maps, including the magnetization field $\mathbf{m}$, demagnetization field $\mathbf{H}_\mathrm{d}$, and strain field $E$, are written either at convergence or after every $N$ steps. The dashed borders indicate the stages that involve CPU–GPU data transfer. These stages are minimum in the workflow highlighting our effort to minimize CPU-GPU data transfer overhead.}
    \label{Fig:flowchart}
\end{figure}

In this section, we introduce the governing equations of our micromagnetics model and their weak forms. We also explain how these equations are numerically implemented in our GPU-resident and tensorized workflow. Throughout this paper, we adopt the following notation convention: vectors and vector fields are denoted in bold (e.g., $\mathbf{m}$), while tensors and tensor fields are written in standard (non-bold) font (e.g., $E$), unless otherwise noted. We summarize the complete workflow of \texttt{CuPyMag}, including data transfers and output handling in Fig.~\ref{Fig:flowchart}.

\subsection{Micromagnetics}
\texttt{CuPyMag} simulates a macroscopic ellipsoid body $\mathcal E\subset \mathbb{R}^3$ containing a cuboid computation domain $\Omega_{\mathrm{c}}\subset\mathcal{E}$, with $V(\Omega_{\mathrm{c}}) \ll V(\mathcal{E})$. Here $V(\cdot)$ denotes the volume. We partition $\Omega_{\mathrm{c}}$ into two domains, namely a defect of arbitrary shape $\Omega_{\mathrm{d}}\subset\Omega_{\mathrm{c}}$ and the ferromagnetic host $\Omega_{\mathrm{m}} =\Omega_{\mathrm{c}}\setminus\Omega_{\mathrm{d}}$. The micromagnetics energy is \cite{brown1963micromagnetics,brown1966magnetoelastic}:
\begin{align}
{\Psi} & =\int_{\mathcal{E}}\Bigg\{\grad\vb{m}:A\grad\vb{m}+\kappa_{1}(\mathrm{m}_{1}^{2}\mathrm{m}_{2}^{2}+\mathrm{m}_{2}^{2}\mathrm{m}_{3}^{2}+\mathrm{m}_{3}^{2}\mathrm{m}_{1}^{2})+\frac{1}{2}[E-E_0(\vb{m})]:\mathbb{C}[E-E_0(\vb{m})]\nonumber\\
&  -\sigma_{\mathrm{ext}}:E-\mu_{0}m_s\vb{H_{\mathrm{ext}}\cdot m} \Bigg\}\mathrm{d\vb{x}} +\int_{\mathbb{R}^{3}}\frac{\mu_{0}}{2}\left|\vb{H}_\mathrm{d}\right|^{2}\mathrm{d\vb{x}}.\label{MicromagneticsEnergy}
 \end{align}
We describe the meaning of each notation in Eq.~\ref{MicromagneticsEnergy} in Appendix A. For the physical meaning of each energy term and other details, we refer the readers to our previous works \cite{balakrishna2021tool,renuka2022design}. Here, $\vb{m} = \vb{M}/m_s$ is the normalized magnetization (i.e., $|\vb{m}|=1$), $E=(\grad\vb{u}+\grad\vb{u}^\mathrm{T})/2$ is the linear strain (derived from displacements $\vb{u}$), and $E_0(\vb{m})$ is the spontaneous strain. The external field $\vb{H}_{\mathrm{ext}}$ is decreased in increments up to a critical coercive field $\vb{H}_{\mathrm{c}}$, at which the magnetization field in the domain reverses. The demagnetization field $\vb{H}_{\mathrm{d}}$ is computed from the demagnetization potential $U$, which is obtained by solving the magnetostatic equilibrium equation:
\begin{equation}
    \grad^2 U = \grad\cdot\vb{M}, ~ \vb{H}_{\mathrm{d}} = -\grad U.
\end{equation}
We use the ellipsoid theorem and decompose the magnetization, demagnetization, and strain fields into two parts \cite{balakrishna2021tool}:
\begin{equation}
    \vb{m} = \vb{\widetilde{m}+\bar{m}},~ \vb{H}_{\mathrm{d}} = \widetilde{\vb{H}}_{\mathrm{d}} + \bar{\vb{H}}_{\mathrm{d}}, ~E = \widetilde{E}+\bar{E}.
\end{equation}
The spatially varying components $\widetilde{\vb{m}}$, $\widetilde{\vb{H}}_{\mathrm{d}}$ and $\widetilde{E}$ represent the local perturbations of the magnetization, demagnetization, and strain fields within the computational domain $\Omega_{\mathrm{c}}$. Their volume integrals vanish inside $\Omega_{\mathrm{c}}$ and are set to zero outside the domain. The homogeneous fields, $\vb{\bar{m}}$, $\vb{\bar{H}}_{\mathrm{d}}$ and $\bar{E}$, represent the volume averages of the corresponding fields in $\Omega_{\mathrm{c}}$. The total demagnetization field is calculated as
\begin{equation}
    \vb{H}_{\mathrm{d}} = \widetilde{\vb{H}}_{\mathrm{d}} + \bar{\vb{H}}_{\mathrm{d}} = -\grad\widetilde{U} - m_sN_{\mathrm{d}}\bar{\vb{m}},
\end{equation}
in which $\widetilde{U}$ is solved as 
\begin{equation}
    \grad^2 \widetilde{U} = m_s\grad\cdot\widetilde{\vb{m}},~\vb{x}\in\Omega_{\mathrm{c}},\label{demag}
\end{equation}
and $N_\mathrm{d}\in\mathbb{R}^{3\times3}$ is the demagnetization factor of the ellipsoid $\mathcal{E}$. Second, we update the strains $E$ by solving for mechanical equilibrium:
\begin{equation}
    \grad\cdot\mathbb{C}[E-E_0(\vb{m})] = \vb{0},~\vb{x}\in\Omega_{\mathrm{c}}. \label{elasticity}
\end{equation}
The strain is similarly decomposed into the homogeneous and heterogeneous parts, and the details could be found in \cite{PhysRevMaterials.9.044407}. Finally, the LLG equation \cite{LandauLifshitz1935,1353448}
\begin{equation}
    \frac{\partial\mathbf{m}}{\partial \tau}=-\mathbf{m}\times\vb{H}_{\mathrm{eff}}-\alpha\mathbf{m}\times(\mathbf{m}\times\vb{H}_{\mathrm{eff}}) \label{Eq:LLG}
\end{equation}
is solved using the GSPM \cite{wang2001gauss}. Here the effective field is given by $\vb{H}_{\mathrm{eff}}=-\frac{1}{\mu_0 m_s^2}\frac{\delta\Psi}{\delta\mathbf{m}}$ and the dimensionless time step is defined as $\tau=\gamma m_s t$. We note that $\gamma$ is the gyromagnetic ratio and $\alpha$ is the damping constant. The GSPM allows for numerical stability comparable to other non-linear implicit methods, while we only need to solve one type of Poisson-type PDE:
\begin{equation}
    (1-\epsilon\grad^2)m = f,~\vb{x}\in\Omega_{\mathrm{m}}, \label{GS}
\end{equation}
in which $\epsilon$ is some small constant that is related to certain model parameters, such as the damping constant $\alpha$, the LLG time step, and the system's characteristic length scale. The first-order derivative of the free energy (Eq.~\ref{MicromagneticsEnergy}) with respect to $\vb{m}$ (excluding the $\grad\vb{m}$ term) is $\vb{f}$. Here we drop the boldface for $\vb{m}$ and $\vb{f}$ as Eq.~\ref{GS} is solved componentwise, and we do not use indices, as Eq.~\ref{GS} represents the general form. In addition, we exclude the defect area to ensure zero magnetization inside the defect and apply Neumann boundary conditions at $\partial\Omega_{\mathrm{d}}$. A typical order of magnitude for $\epsilon$ in our previous calculations is $10^{-2}$ \cite{balakrishna2021tool,ARB2021Permalloy,renuka2022design,PhysRevMaterials.9.044407}. Therefore, the solution $m$ is close to $f$, and thus Eq.~\ref{GS} could be quickly solved with iterative solvers. As a result, the GSPM previously implemented on the CPUs with FDM \cite{wang2001gauss} could also potentially be an efficient algorithm on GPUs with FEM. The details of the GSPM are attached in Appendix B and could be found in \cite{wang2001gauss,zhang2005phase,balakrishna2021tool,renuka2022design}. In summary, Eqs.~\ref{demag},\ref{elasticity},\ref{GS} are the strong forms of the PDEs to solve.

\subsection{Weak forms}
To solve these equations using FEM, we employ variational methods to obtain their weak forms. The derivations of Eqs.~\ref{demag},\ref{elasticity},\ref{GS} follow standard FEM procedures (see e.g., \cite{10.1063/5.0105613}), and in particular the derivation of Eq.~\ref{GS} is analogous to that of the Poisson equation. For brevity, we summarize the final results here. Let $\mathcal{V}$ be the Sobolev space $H^1(\Omega_\mathrm{c})$ subject to periodic boundary conditions. Eqs.~\ref{demag},\ref{elasticity},\ref{GS} are equivalent to the following weak forms:
\begin{enumerate}
    \item[(a)] Magnetostatic equilibrium (Eq.~\ref{demag}): Find $\widetilde{U}\in \mathcal{V}$ such that
    \begin{equation}
        \int_{\Omega_{\mathrm{c}}} \grad v\cdot \grad \widetilde{U} \dd\vb{x} = -m_s\int_{\Omega_{\mathrm{c}}} v (\grad\cdot\widetilde{\vb{m}}) \dd\vb{x}, ~\forall v\in \mathcal{V}. \label{weak_demag}
    \end{equation}
    \item[(b)] Mechanical equilibrium (Eq.~\ref{elasticity}): Find $\vb{u}\in \mathcal{V}^3$ such that
    \begin{equation}
        \int_{\Omega_{\mathrm{c}}} E(\vb{v}): \mathbb{C} E(\vb{u})\dd\vb{x} = \int_{\Omega_{\mathrm{c}}} E(\vb{v}): \mathbb{C}E_0(\vb{m})\dd\vb{x},~\forall \vb{v}\in\mathcal{V}^3, \label{weak_elasticity}
    \end{equation}
    where $E(\vb{u}) := (\grad\vb{u}+\grad\vb{u}^{\operatorname{T}})/2$.
    \item[(c)] The GSPM related PDE: Find $m\in \mathcal{V}$ such that
    \begin{equation}
        \int_{\Omega_{\mathrm{c}}} (mv+\epsilon\grad m\cdot \grad v)\dd\vb{x} = \int_{\Omega_{\mathrm{c}}} fv \dd\vb{x},~\forall v\in\mathcal{V}. \label{weak_GS}
    \end{equation}
    
\end{enumerate}

\subsection{Stiffness matrix assembly and RHS assembly}
\texttt{CuPyMag} performs global matrix assembly only at the beginning and transfers all of the data to the GPU for subsequent linear solves, right-hand side (RHS) assembly and spatial derivatives. For completeness, we list the matrix assembly details: Suppose the variational space $\mathcal{V}$ is approximated by the finite-dimensional subspace $\mathcal{V}_h = \operatorname{span}\{\phi_1,...,\phi_N\}$, where $\phi_i$ is the shape function on the $i$-th node and $N$ is the total number of nodes. Throughout this subsection, we use $i,j$ to denote nodal indices, and $a,b,r,s$ to denote Cartesian components. Grouped indices such as $(k,rs)$ combine a nodal index $k$ with tensor components $r,s$. With this notation rule, the left-hand side (LHS) stiffness matrices $K$ for Eqs.~\ref{weak_demag},\ref{weak_elasticity},\ref{weak_GS} are, respectively:
\begin{equation}
    \begin{aligned}
        &  (K)_{ij}^{\text{ms}} = \int_{\Omega_{\mathrm{c}}}\grad \phi_i\cdot \grad \phi_j \dd\vb{x}, \\
        &  (K)_{(i,a),(j,b)}^{\text{mech}} = \int_{\Omega_{\mathrm{c}}}E(\phi_j\hat{\vb{e}}_b):\mathbb{C}E(\phi_i\hat{\vb{e}}_a) \dd\vb{x}, \\
        &  (K)_{ij}^{\text{GS}} = \int_{\Omega_{\mathrm{c}}}\qty(\epsilon\grad \phi_i\cdot \grad \phi_j + \phi_i\phi_j)\dd\vb{x}.
    \end{aligned} \label{LHS}
\end{equation}
Here $\hat{\vb{e}}_a,\hat{\vb{e}}_b$ are unit vectors in $a,b$ directions, respectively. In addition, \texttt{CuPyMag} also pre-assembles the RHS mass matrices $F$, allowing the RHS vector to be computed efficiently at each time step via a single matrix–vector multiplication with the current field vector. Since the RHS must be updated at every time step in micromagnetic simulations, this approach significantly reduces computational overhead, particularly on GPUs. The mass matrices assembled as
\begin{equation}
        \begin{aligned}
        & (F)_{ij,a}^{\text{ms}} = -\int_{\Omega_{\mathrm{c}}}\phi_i(\grad \phi_j)_a  \dd\vb{x},  \\
        &  (F)_{(i,a),(j,rs)}^{\text{mech}} = \dfrac{1}{2}\int_{\Omega_{\mathrm{c}}}E(\phi_i\vb{\hat{e}}_a):\mathbb{C}\phi_j(\hat{\vb{e}}_r\otimes\hat{\vb{e}}_s+\hat{\vb{e}}_s\otimes\hat{\vb{e}}_r) \dd\vb{x}, \\
        &  (F)^{\text{GS}}_{ij} = \int_{\Omega_{\mathrm{c}}} \phi_i\phi_j\dd\vb{x}.
    \end{aligned} \label{RHS_F}
\end{equation}
Finally, the RHS vectors are assembled as (Einstein's summation rule assumed):
\begin{equation}
    \begin{aligned}
        & \qty(\vb{b}^{\text{ms}}_h)_i = m_s F^{\text{ms}}_{ij,a} \qty(\vb{m}_h)_{j,a}, \\
        & \qty(\vb{b}^{\text{mech}}_h)_{(i,a)} = (F)_{(i,a),(j,rs)}^{\text{mech}}\qty(E_{0}(\vb{m}_h))_{(j,rs)}, \\
        & \qty(\vb{b}^{\text{GS}}_h)_i = F^{\text{GS}}_{ij}\qty(\vb{f}_h)_j.
    \end{aligned} \label{RHS_b}  
\end{equation}
In Eq.~\ref{RHS_b}, we use the subscript $h$ to indicate that a field is discretized in $[\mathcal{V}_h]^d$, in which $d$ is determined by whether it is a scalar, vector, or a tensor field. Noticeably, the matrices $(F)_{a}^{\text{ms}}$ represent the minus of the gradient operator $-(\grad)_a$ under periodic boundary conditions. \texttt{CuPyMag} reuses these matrices to compute spatial derivatives:
\begin{equation}
\begin{aligned}
    (\vb{H}_{\mathrm{d},h})_a &= -(\grad)_a \vb{U}_h = (F)^{\text{ms}}_a \vb{U}_h, \\
    (E_h)_{a,b} &= \dfrac{1}{2}\qty[(\grad)_a (\vb{u}_h)_b + (\grad)_b (\vb{u}_h)_a] = -\dfrac{1}{2}\qty[(F)^{\text{ms}}_a (\vb{u}_h)_b + (F)^{\text{ms}}_b (\vb{u}_h)_a].
\end{aligned}
\end{equation}
As a result, all the derivatives are performed via matrix-vector multiplications on the GPU.

\subsection{Volume average}
Implementing the ellipsoid theorem in computing the evolving fields requires frequent volume averages at each time step:
\begin{equation}
    \bar{g} = \dfrac{1}{V(\Omega_{\mathrm{c}})}\int_{\Omega_{\mathrm{c}}}g(\vb{x})\dd \vb{x}
\end{equation}
Here, $g$ represents an arbitrary field. On an unstructured finite-element mesh \texttt{CuPyMag} approximates the integral with an element-wise Gauss quadrature sum:
\begin{equation}
    \bar{g} \approx \dfrac{\sum\limits_{e=1}^{N_e}\sum\limits_{p=1}^{N_{gp}}g_e(\vb{r}_p)|J(e,p)|w_p}{\sum\limits_{e=1}^{N_e}\sum\limits_{p=1}^{N_{gp}}|J(e,p)|w_p},
\end{equation}
where $w_p$ and $\vb{r}_p$ represent the weights and the coordinates of the Gauss points and $J(e,p)$ is the Jacobian between physical and reference coordinates at the $p$-th Gauss point in the $e$-th element. $N_e$ represents the total number of elements and $N_{gp}$ is the number of Gauss points in each element.

To avoid redundant computation of recurring quantities, \texttt{CuPyMag} precomputes and stores the following quantities on the GPU:
\begin{enumerate}
    \item Determinant of Jacobians $|J(e,p)|$ at each Gauss point $p$ in each element $e$.
    \item Shape functions $\phi_i(\vb{r}_p)$ for each node $i$ and Gauss point $p$.
\end{enumerate}
In the next step, the volume average of a field $g\in\mathcal{V}_h^{n}$ is calculated as follows:
\begin{enumerate}
    \item $g_{\mathrm{node}}$ (shape $(N_e, N_i, n$) stores the coalesced nodal values at each node in the element connection map (shape $(N_e, N_i)$). Here, $N_i$ is the number of nodes per element.
    \item $g_e(\vb{r}_p) = \sum_i \phi_i(\vb{r}_p)g_{\mathrm{node}}[e,i,c]$ (shape $(N_e, N_{gp}, n$) sums over nodes $i$ and represents the value of $g$ at each component $c$ and at each Gauss point $p$ and element $e$. It is efficiently implemented by \texttt{cupy.einsum}.
    \item The sum $\sum_{e=1}^{N_e}\sum_{p=1}^{N_{gp}}g_e(\vb{r}_p)|J(e,p)|w_p$ and  $\sum_{e=1}^{N_e}\sum_{p=1}^{N_{gp}}|J(e,p)|w_p$ are respectively calculated by CuPy's element-wise summation. 
\end{enumerate}
While volume averaging with Gauss quadrature on GPUs is well studied in general finite element methods \cite{MACIOL20101093,BANAS20141319,CAMES,10.1145/2427023.2427027}, to the best of our knowledge, no micromagnetic FEM work has reported an analogous GPU-resident algorithm for performing repeated volume averages. Therefore, this module in \texttt{CuPyMag} serves as an innovative, fully batched FEM volume-average kernel for micromagnetic simulations with open-source implementations.

Overall, once the system is assembled on the CPU, the data resides on the GPU for the subsequent operations, such as spatial derivatives, volume averages, right-hand side updates, and iterative solvers. All of these operations are carried out exclusively using CuPy's BLAS-accelerated tensor operations \cite{10.1145/3431921,cupy_learningsys2017}, ensuring high efficiency throughout the simulation. 

\subsection{Boundary conditions and solvers}
\texttt{CuPyMag} uses periodic boundary conditions to represent an infinite array of periodically arranged defects. This periodic boundary condition is enforced by identifying the corresponding degrees of freedom (DOFs) on matching nodes at opposite faces of the domain. In practice, \texttt{CuPyMag} enforces periodic boundary conditions by “wrapping” any node whose coordinate in a given direction lies at the domain’s maximum extent to the corresponding node at the domain's minimum extent in that direction. Then \texttt{CuPyMag} inserts each wrapped node pair into a hash map and assigns a unique global index. By doing so, the two nodes in the pair are treated as one unified node. This approach offers a straightforward method for implementing periodicity through node pairings. However, its major limitation is that it only works for rectangular domains with perfectly matched node distributions at opposite boundaries. For domains with irregular shapes or mismatched meshes, additional procedures such as face interpolation or tree-based nearest-neighbor searches may be required.

We note that all the stiffness matrices,
\begin{equation}
    K\in \{K^{\mathrm{ms}},K^{\text{mech}},K^{\text{GS}} \},
\end{equation}
are symmetric by construction and satisfy the semi‑definiteness property,
\begin{equation}
    \vb{v}^{\operatorname{T}}K\vb{v}\geq 0, \forall \vb{v}\in \mathbb{R}^N,K\in \{K^{\mathrm{ms}},K^{\text{mech}},K^{\text{GS}} \}. 
\end{equation}
Under periodic boundary conditions, the nullspace of $K$ only corresponds to constant fields:
\begin{equation}
    \operatorname{dim(ker}(K)) = d,
\end{equation}
where $d$ is the number of field components. All other vectors outside the nullspace of $K$ have a strictly positive relation: $\vb{v}^{\operatorname{T}}K\vb{v} > 0$. Therefore, it suffices to pin $d$ DOFs to 0, so that the reduced matrix $\widetilde{K}$ is strictly symmetric positive‐definite. Intuitively, this eliminates the global constant of demagnetization potential or strain displacements, and thus a unique solution is obtained. Therefore, the Conjugate Gradient (CG) solver applies to all of the equations in \texttt{CuPyMag} directly and is guaranteed to converge.

\section{Performance Benchmark}
\begin{figure}[!t]
    \centering
    \includegraphics[width=\linewidth]{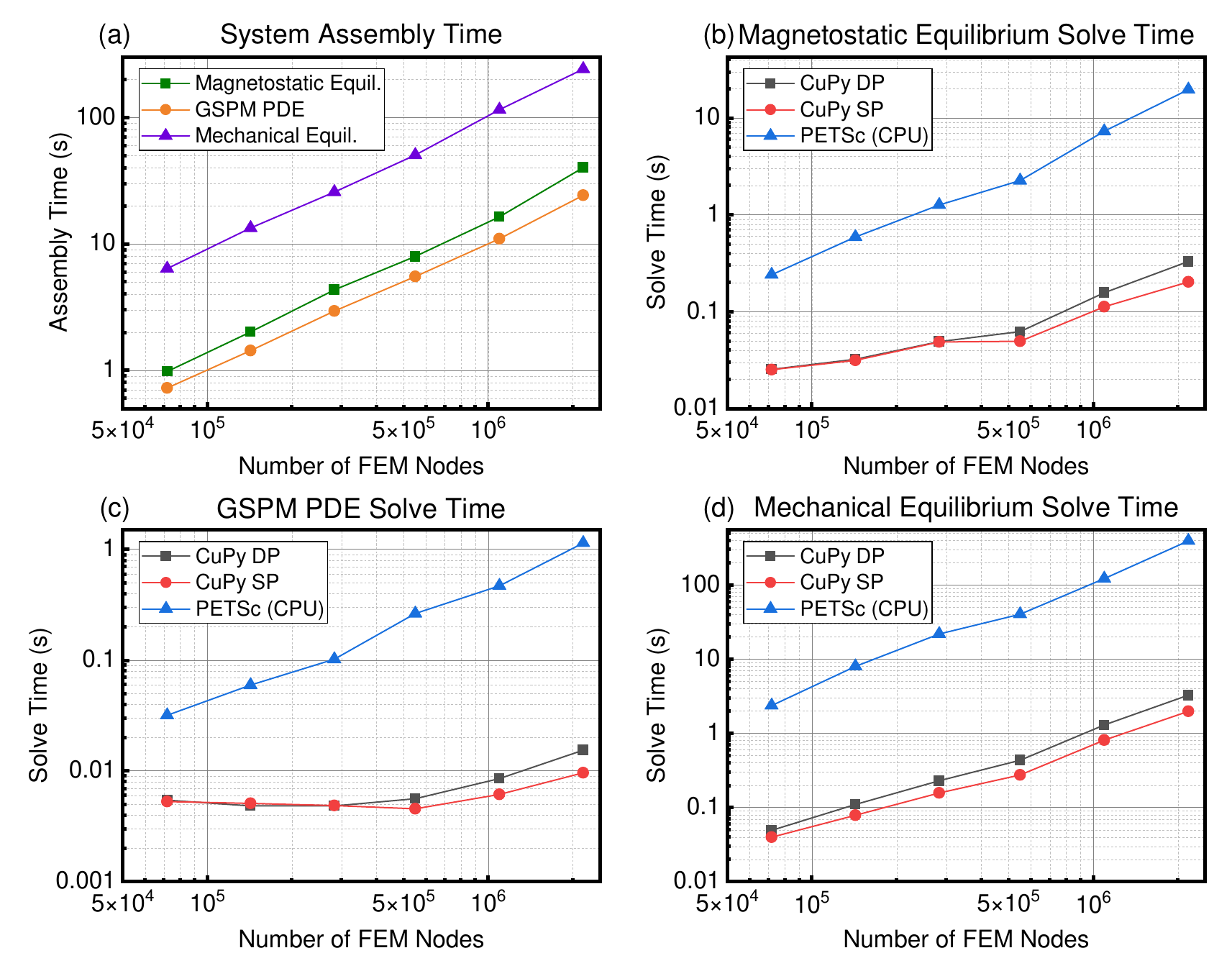}
    \caption{Benchmark of system assembly and linear solver performance for a linear hexahedral element system. (a) Assembly time for the magnetostatic equilibrium system, the Gauss–Seidel projection method (GSPM) system, and the mechanical equilibrium system (represented in a distinct color scheme, see online version). (b–d) Conjugate gradient (CG) solve time for these three systems with different precisions and backend implementations. For each system, we test the solve time using CuPy in double precision (DP), CuPy in single precision (SP), and using PETSc on CPUs (in DP and via \texttt{petsc4py} \cite{DALCIN20111124,osti_2565610}). We do not use a preconditioner in any of the CG solvers.}
    \label{Fig:benchmark}
\end{figure}

\begin{figure}[!t]
    \centering
    \includegraphics[width=\linewidth]{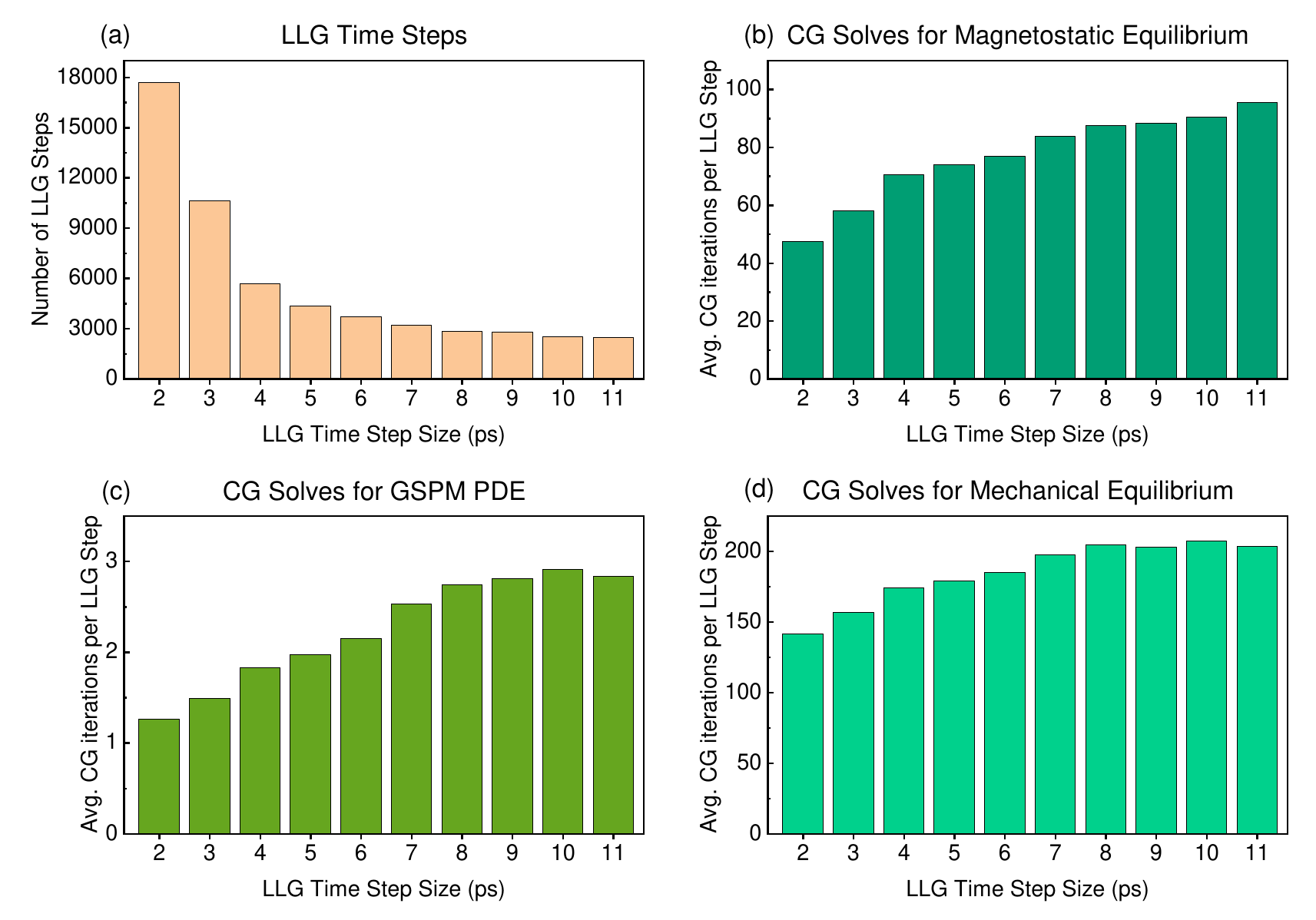}
    \caption{Benchmark for an example micromagnetic calculation with 550k nodes on a linear hexahedral mesh. The computational domain contains a non-magnetic defect at its geometric center and is initialized with uniform magnetization. The LLG equations are solved by the Gauss-Seidel projection method (GSPM) iteratively until convergence, with LLG time steps ranging from 2 ps to 11 ps. (a) The total number of LLG steps required to converge for each calculation. (b-d) The average number of CG iterations for solving the magnetostatic equilibrium, GSPM, and mechanical equilibrium systems. These are grouped with a consistent color family to highlight their conceptual relation. All calculations are performed in double precision.}
    \label{Fig:GSPM}
\end{figure}

In this section, we benchmark \texttt{CuPyMag}'s performance on both linear hexahedral meshes (Figs.~\ref{Fig:benchmark},\ref{Fig:GSPM}) and unstructured linear-hexahedral meshes (Fig.~\ref{Fig:TotalBenchmark}). Across all problem sizes and different meshes, runtime grows linearly or even sub-linearly. Such scaling, which indicates high GPU resource utilization, exists up to the largest problem that approaches the GPU memory limit. Therefore, \texttt{CuPyMag} is memory-bound rather than limited by compute throughput. The vectorized CG solve on GPU significantly outperforms the CPU, and since \texttt{CuPyMag}’s core operations are tensorized and parallel on the GPU, we expect this performance gain to extend to the overall workflow. In a representative micromagnetics problem, \texttt{CuPyMag} remains stable and efficient with time steps of 5--10 ps, which we attribute to the implementation of GSPM. Collectively, these results demonstrate \texttt{CuPyMag}'s suitability in solving large-scale micromagnetics problems within an FEM framework. All the GPU calculations presented in this section use CUDA version 12.6.

\subsection{Assembly and solver scaling}
Fig.~\ref{Fig:benchmark}(a) shows the system assembly time for the three governing equations, the magnetostatic equilibrium (Eq.~\ref{weak_demag}), the GSPM-related PDE (Eq.\ref{weak_GS}), and the mechanical equilibrium (Eq.~\ref{weak_elasticity}), solved with a structured linear-hexahedral mesh. Each computation is allocated 16 CPU threads on an AMD EPYC 7763 CPU. The assembly time for all three systems scales linearly with the problem size up to $>2$M nodes. This linear scaling suggests that Numba’s JIT-compiled routines achieve constant per-DOF assembly cost while avoiding the large overheads of pure Python, such as the Global Interpreter Lock (GIL) \cite{beazley2010gil}. As a result, the matrix assembly is no longer a computational bottleneck in micromagnetic calculations.

Fig.~\ref{Fig:benchmark}(a) also shows the differences in absolute assembly cost across the three systems due to the complexity of their weak forms and the number of unknowns per node. The mechanical equilibrium system assembly is the most time-consuming for a given mesh because it involves a vector field (3 displacement components per node), leading to larger system matrices. This system also involves more complex integrands, as represented in Eqs.~\ref{LHS},\ref{RHS_F}. By contrast, the magnetostatic equilibrium and the GSPM systems are faster to assemble because of their similarity to a Poisson equation. Nonetheless, even a large system with $>2$M nodes could be constructed quickly since their assembly time is a few hundred seconds (see Fig.~\ref{Fig:benchmark}(a)). Once constructed, all tensors are uploaded to the GPU for subsequent tensorized operations.

Figs.~\ref{Fig:benchmark}(b-d) report the CG solver runtimes for the three linear systems (magnetostatic equilibrium, GSPM-related PDE, and mechanical equilibrium), comparing the GPU implementations in double and single precision and CPU implementation in double precision. The CPU solver is represented with PETSc (via \texttt{petsc4py} \cite{DALCIN20111124,osti_2565610}) using 16 CPU threads on 2.4 GHz AMD EPYC 7763. For GPU computations, we use CuPy's native CG solver implemented in \texttt{cupyx.scipy.sparse.linalg.cg} \cite{cupy_learningsys2017} on a single NVIDIA A100 GPU. For each system, the solve time is averaged over 50 runs of solving $A\vb{x=b}$. A convergence tolerance of $\frac{\|\vb{b}-A\vb{x}\|_2}{\|\vb{b}\|_2} < 1\times 10^{-7}$ is set for all calculations.

Note that in Figs.~\ref{Fig:benchmark}(b-d), we do not use preconditioners in the CG solvers. This choice is made because CuPy’s CG routine only supports explicit preconditioners of the form $M\approx K^{-1}$, where $M$ is the preconditioning matrix. For solving elliptic partial differential equations, such as the computationally expensive equations in our model Eq.~\ref{demag},\ref{elasticity}, $K^{-1}$ is dense because of the global coupling of the magnetic and elastic fields. This makes any sparse approximation of $M$ ineffective. Although algebraic multigrid is considered to be effective for FEM problems, implementing an efficient GPU version within CuPy is nontrivial. Additionally, as we will discuss in subsection \ref{3.3}, \texttt{CuPyMag}'s performance is memory-bound rather than iteration-bound. For these reasons, all benchmarks are performed without using preconditioners in CG solver.

The key result in Figs.~\ref{Fig:benchmark}(b-d) is that GPU solvers are 1--2 orders of magnitude faster than their CPU counterparts. Since CG is an $O(N)$ algorithm, the solve time is expected to scale linearly with problem size. However, in Figs.~\ref{Fig:benchmark}(b-c), magnetostatic equilibrium and GSPM PDE systems exhibit nearly constant or sub-linear scaling, suggesting that GPU resources are underutilized and the runtime for small systems is dominated by overheads such as memory allocation and kernel compilation. As the problem size increases, more threads are engaged, leading to better hardware saturation. This ``small system'' behavior exists up to 500k nodes, see Figs.~\ref{Fig:benchmark}(b,c), indicating that a problem size of up to half a million is not large enough to saturate an A100. These features highlight the potential of $\texttt{CuPyMag}$ for addressing large-scale problems.

Comparing the solve time across the three systems, we note that the GSPM PDE is much faster than the magnetostatic equilibrium system, a system with the same linear operator as the standard Poisson equation. This is because the presence of the mass term in $K^{\text{GS}}$ (Eq.~\ref{LHS}) makes the GSPM PDE better conditioned. We will show in the next subsection that it reduces the number of iterations required to converge. By contrast, the mechanical equilibrium system is the most expensive to compute, with its solve time scaling linearly in all the test cases, see Fig.~\ref{Fig:benchmark}(d).

We next compare the single and double-precision GPU calculations, as shown in Fig.~\ref{Fig:benchmark}(b-d). As long as the problem is large enough to saturate the GPU, single precision provides nearly a two-fold speedup for large problem sizes. This is achieved primarily by halving the memory bandwidth required for vector and matrix operations. Therefore, if a lower precision data format is sufficient for a micromagnetics study, simulations with a single precision format can achieve almost twice the throughput within the same GPU memory limit.

\subsection{Time-integration investigation}
In this subsection, we benchmark the time‐integration performance with GSPM using a test problem. We model a domain with a structured mesh containing about 550k FEM nodes and a non-ferromagnetic cuboid defect in the center. The material parameters are set to be the same as those of the soft magnet Ni$_{70}$Fe$_{30}$ \cite{ARB2021Permalloy,RevModPhys.25.42}. We initialize the domain $\Omega_{\mathrm{m}}$ with uniform magnetization and under a fixed external field of 200 A/m. The LLG equation is solved iteratively by GSPM until the average per-DOF L2 norm of the magnetization update satisfies the condition $\expval{\|\vb{m}^n-\vb{m}^{n-1}\|_2}_{\mathrm{DOF}}<5\times 10^{-10}$. The Gilbert damping constant is set to $\alpha=0.15$ for all the calculations. We summarize the results in Fig.~\ref{Fig:GSPM}.

Fig.~\ref{Fig:GSPM}(a) shows a monotonic decrease of the total LLG steps $N_{\mathrm{LLG}}$ as the time step $\Delta t$ increases from 2 ps to 11 ps. A key highlight is that the number of LLG steps to converge, $N_{\mathrm{LLG}}$ rapidly decreases from $\sim 18,000$ steps at $\Delta t = 2$ ps to a third of this value $\sim 6,000$ steps at $\Delta t = 4$ ps. This demonstrates the numerical stability of the GSPM compared to other explicit methods that often lead to a $\Delta t$ constraint at sub-pico seconds \cite{wang2001gauss,LI2023107512}. In addition, without any preconditioning, each GSPM PDE requires only $\bar{N}_{\mathrm{CG}}^{\mathrm{GSPM}}=1-3$ CG iterations as shown in Fig.~\ref{Fig:GSPM}(c). 

Fig.~\ref{Fig:GSPM}(b, d) shows the number of CG iterations required to solve the magnetostatic and mechanical equilibrium equations, respectively. The magnetostatic equilibrium equation requires $\bar{N}_{\mathrm{CG}}^{\mathrm{ms}}=$40-90 CG iterations and the mechanical equilibrium equation requires $\bar{N}_{\mathrm{CG}}^{\mathrm{mech}}=$150-200 CG iterations to converge, respectively. This relatively larger number of CG iterations (in comparison to the GSPM PDE in Fig.~\ref{Fig:GSPM}(c)) explain the increased solve time results for these systems in Figs.~\ref{Fig:benchmark}(b-d). In addition, as $\Delta t$ increases from 2 ps to 11 ps, while $N_{\mathrm{LLG}}$ decreases 6-fold, the average CG iterations $\bar{N}_{\mathrm{CG}}$ for the most computationally expensive equation: mechanical equilibrium only increase around 1/3 as shown in Fig.~\ref{Fig:GSPM}(d). Since the total simulation time scales as $T_{\mathrm{total}}\propto N_{\mathrm{LLG}}\bar{N}_{\mathrm{CG}}$, \texttt{CuPyMag} can efficiently handle time steps of 5--10 ps with GSPM that is larger than a common micromagnetic simulation time step \cite{wang2001gauss,LI2023107512}.

\begin{figure}[!t]
    \centering
    \includegraphics[width=\linewidth]{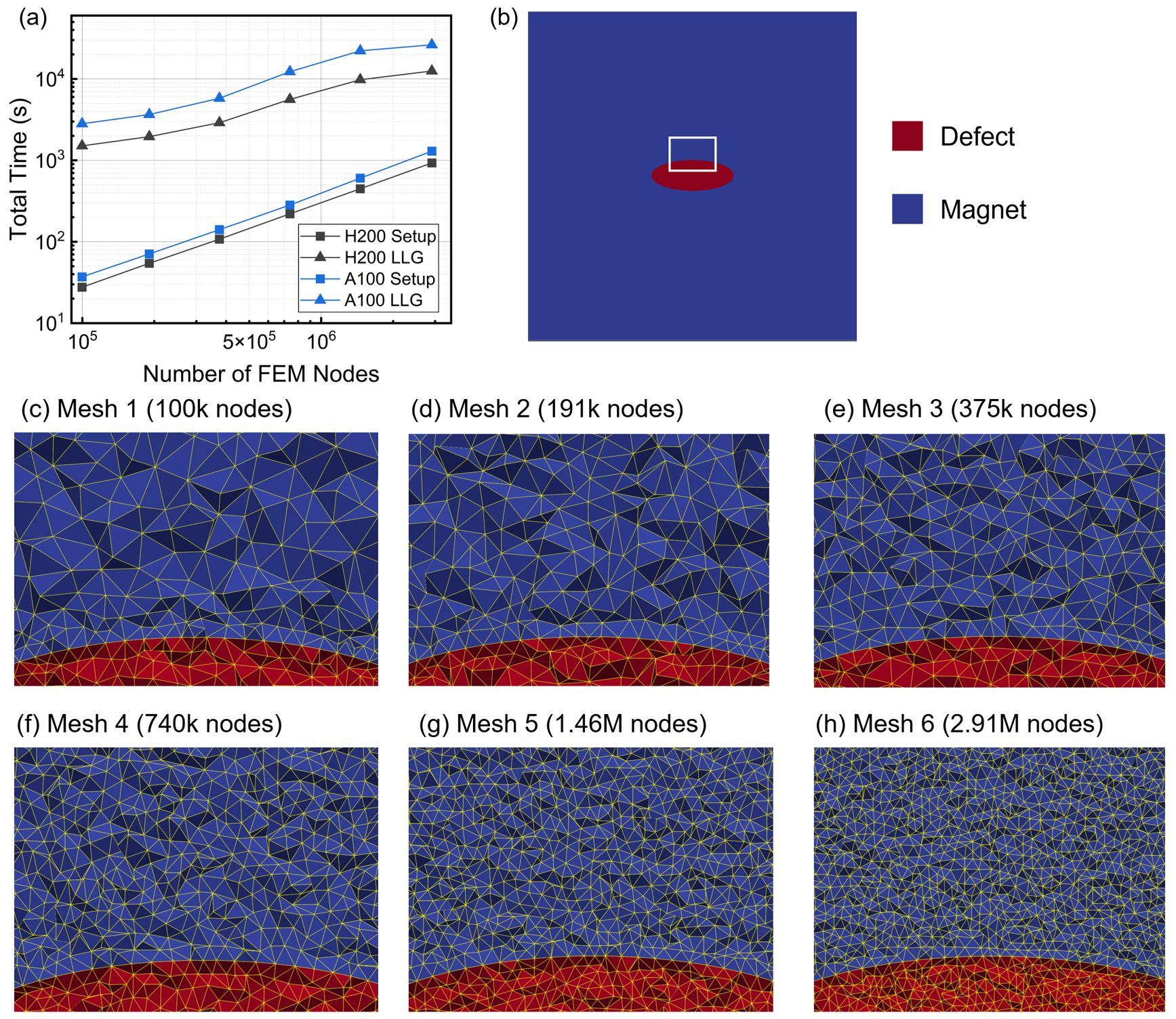}
    \caption{Total runtime benchmark for the full FEM micromagnetics simulation of the hysteresis process of Ni$_{70}$Fe$_{30}$, from $\vb{H}_{\mathrm{ext}}=200~\mathrm{A/m}~\vb{\hat{e}}_1$ to $\vb{H}_{\mathrm{ext}}=-|\vb{H}_{\mathrm{c}}|~\vb{\hat{e}}_1$. (a) Log–log plot of total runtime versus number of FEM nodes, comparing Delta’s H200 node (gray markers) and A100 node (blue markers), with squares denoting system assembly time and triangles denoting the subsequent simulations at all the LLG timesteps. (b) Projection of the sample domain onto the $x-y$ plane. The blue region is the magnetic region, the red ellipse is the nonmagnetic defect, and the white box indicates the subregion used for mesh visualization. (c-h) Linear tetrahedral FEM mesh visualization within the white box denoted in (b) for six increasing mesh densities. All the calculations are performed in double precision.}
    \label{Fig:TotalBenchmark}
\end{figure}

\subsection{Overall runtime test}\label{3.3}
To complement the structured-mesh studies above, we next compute the total runtime for the same representative micromagnetics problem on an unstructured linear tetrahedral mesh. This latter type of finite element mesh can simulate non-trivial defect geometries often encountered in real-world materials. For our calculations, we set the magnetic material parameters to correspond to that of Ni$_{70}$Fe$_{30}$ \cite{ARB2021Permalloy,RevModPhys.25.42}. The stiffness and mass matrices are assembled on the CPU and transferred to the GPU, and the Jacobian determinants and shape functions are precomputed for volume averages. We initialize our computational domain $\Omega_{\mathrm{m}}$ with uniform magnetization under an external field $\vb{H}_{\mathrm{ext}}=200~\mathrm{A/m}~\vb{\hat{e}}_1$. Then \texttt{CuPyMag} follows the workflow described in Fig.~\ref{Fig:flowchart}, with a time step $\Delta t = 3$~ps and a convergence threshold $\expval{\|\vb{m}^n-\vb{m}^{n-1}\|_2}_{\mathrm{DOF}}<5\times 10^{-10}$. The external field is reduced in intervals of $\Delta H_{\mathrm{ext1}}=-20$~A/m. 

Figs.~\ref{Fig:TotalBenchmark}(c-h) show the six meshes of increasing density used in our benchmark tests. These meshes are generated by COMSOL \cite{comsol}, with the same geometry as used in the computations of Fig.~\ref{Fig:benchmark}. Due to space limitations, Figs.~\ref{Fig:TotalBenchmark}(c-h) only display the meshes inside the white box in Fig.~\ref{Fig:TotalBenchmark}(b). A complete visualization of the most dense mesh (Fig.~\ref{Fig:TotalBenchmark}(h)) is illustrated in Appendix C. 

Fig.~\ref{Fig:TotalBenchmark}(a) shows the system setup time and the subsequent runtime (LLG iterations) for each of the six meshes, using one A100 GPU (2.45 GHz AMD EPYC 7763 CPU) and one H200 GPU (2.10 GHz Intel Xeon Platinum 8558 CPU) on the NCSA Delta supercomputer \cite{delta_nsca}, with 12 CPU threads allocated for each calculation. In each case, the setup stage remains a couple of orders of magnitude faster than the coupled LLG time-integration loop (including the magnetostatic and mechanical equilibrium sovlers). The system assembly time scales linearly with the problem size, and the coupled LLG loops simulations scale sub-linearly, indicating efficient GPU utilization. On a single A100 GPU (40 GB memory), \texttt{CuPyMag} scales efficiently to systems up to 3M nodes, completing micromagnetics calculations in under 7 hours. Here it is important to note that by increasing the system size from 100k to 3M nodes (a 30$\times$ increase) results in only a 9$\times$ increase in runtime. This sublinear scaling highlights the slow growth of computational time with system size and indicates that \texttt{CuPyMag} is primarily memory-bound rather than compute-bound. As a result, \texttt{CuPyMag} is particularly well suited for large-scale parameter space sweep studies and high-throughput simulations.

Fig.~\ref{Fig:TotalBenchmark}(a) also compares the total time for micromagnetics calculations on two different GPU hardware. Here, the H200 consistently outperforms the A100 by a factor of 2-3, with greater accelerations seen in larger problem sizes. Although modern GPUs are often considered to optimize for low-precision tensor operations, the Hopper-based H200 \cite{nvidia2024h200} offers very high double-precision throughput, making it well-suited to high-precision scientific simulations. For the largest test case ($\sim$3M nodes), the complete hysteresis simulation takes 3 hours on a single H200 GPU. These results highlight both the high efficiency of \texttt{CuPyMag} and the significant performance improvements offered by new GPU architectures.

\section{Example Calculations}

\begin{figure}[!t]
    \centering
    \includegraphics[width=\linewidth]{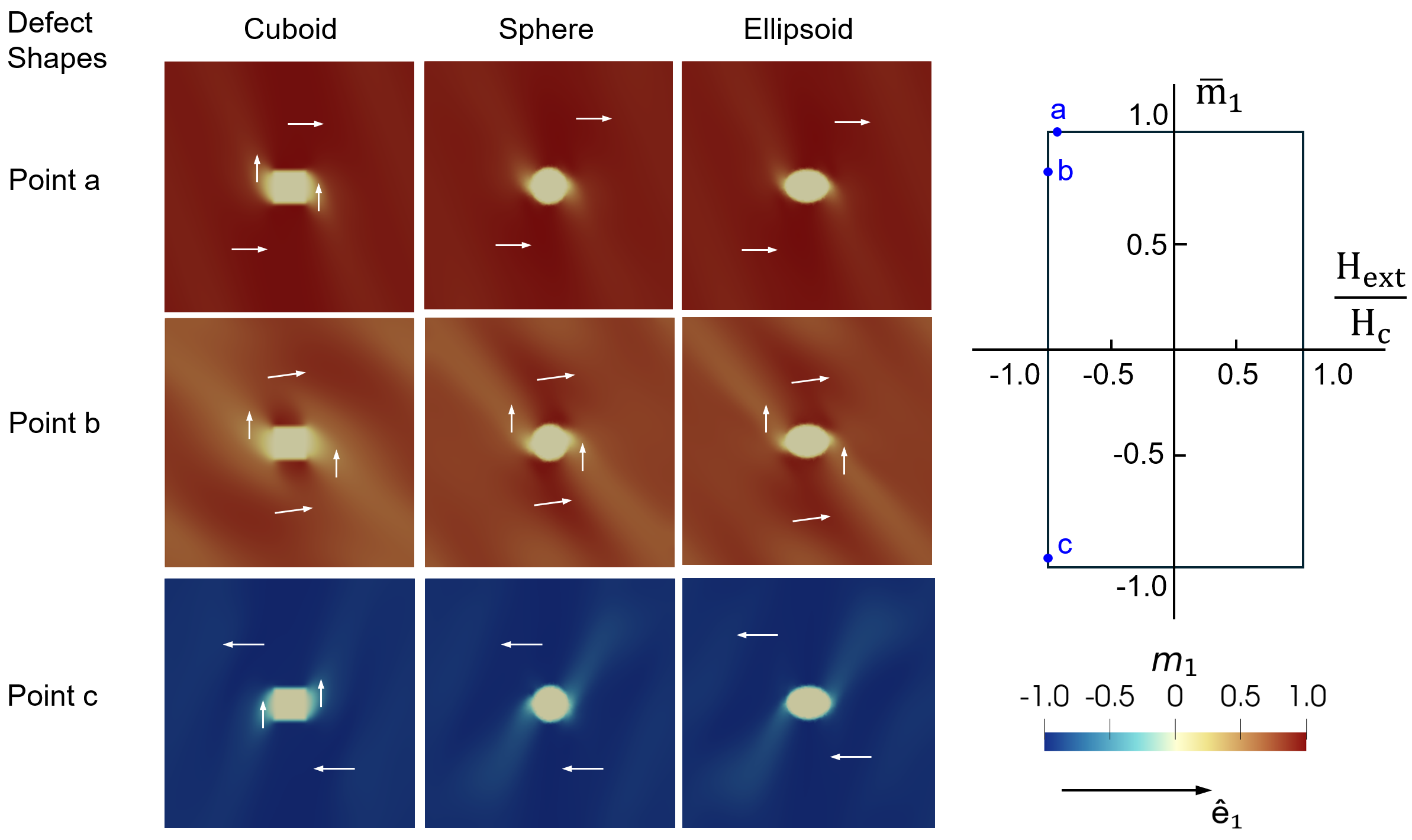}
    \caption{The evolution of needle and stripe-shaped magnetic domains (as 2D projection) around a cuboid, sphere, or ellipsoid-shaped defect, at the points a, b, and c on the hysteresis loop at right. Note that because the $x$‑axis is a normalized field scale, the loops are not identical absolute-field responses but are shown here to highlight the characteristic hysteresis behavior for each defect. The color map shows the magnetization along the $\hat{\vb{e}}_1$ direction: $m_1=\vb{m}\cdot\hat{\vb{e}}_1$.}
    \label{Fig:domains}
\end{figure}

\begin{figure}[!t]
    \centering
    \includegraphics[width=\linewidth]{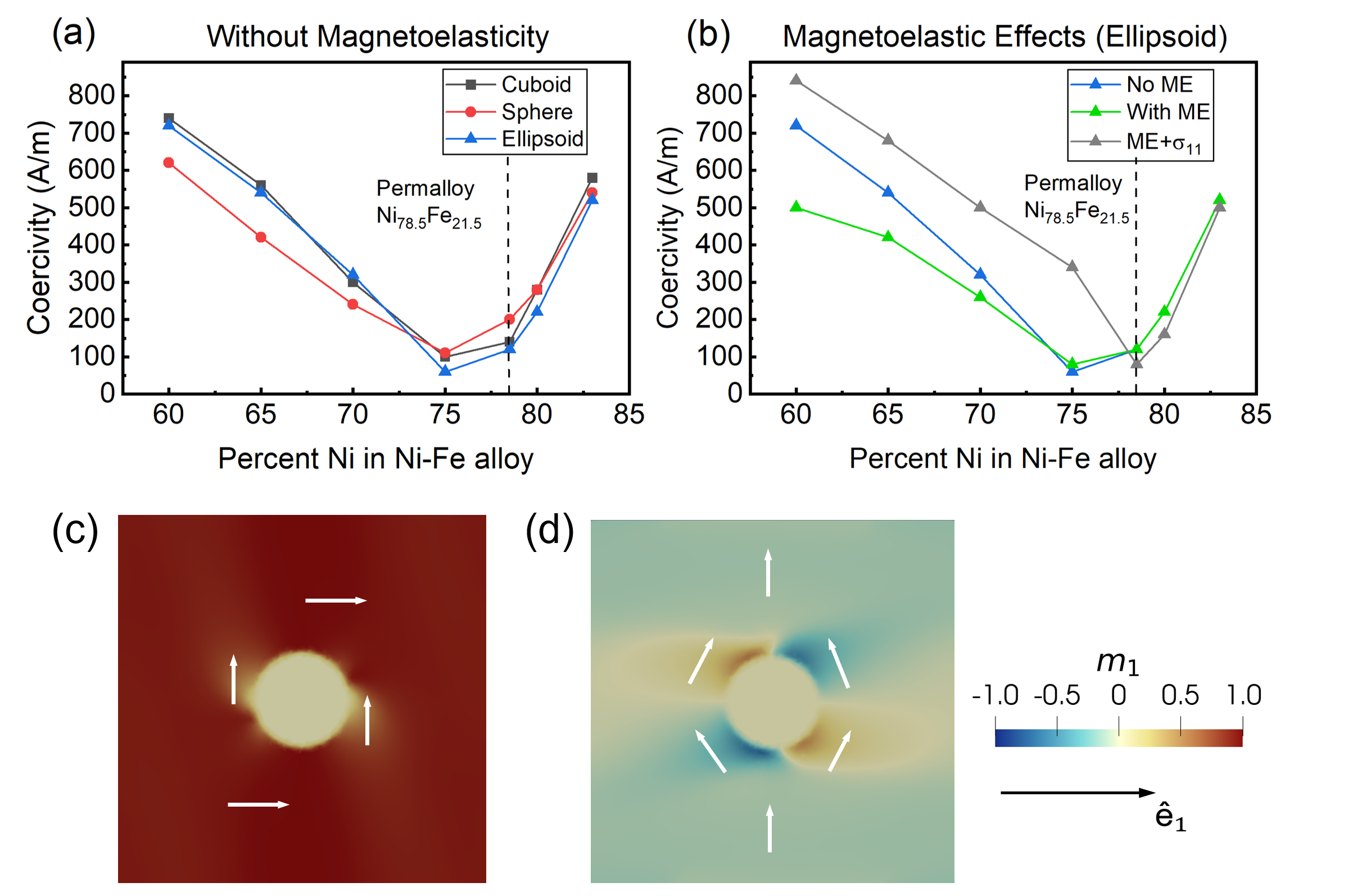}
    \caption{(a) The coercivity of the $\mathrm{Ni}_{x}\mathrm{Fe}_{1-x}$ alloy, with $x$ ranging from 0.60 to 0.83 for cuboid, spherical and ellipsoid shaped defect without considering magnetoelastic (ME) coupling. (b) The coercivity of the $\mathrm{Ni}_{x}\mathrm{Fe}_{1-x}$ alloy, with $x$ ranging from 0.60 to 0.83 for ellipsoid defect, without ME coupling, with ME coupling, and with an uniaxial stress $\sigma_{\mathrm{ext}}=\sigma_{11}\hat{\vb{e}}_1\otimes\hat{\vb{e}}_1 $, $\sigma_{11}=6$ MPa. (c) The magnetization of Galfenol (Fe$_{87}$Ga$_{13}$) \cite{10.1063/1.3618684} without external stress. (d) The magnetization of Galfenol (Fe$_{87}$Ga$_{13}$) with an uniaxial stress $\sigma_{11}=-8$ MPa.}
    \label{Fig:permalloy_galfenol}
\end{figure}

In this section, we present the results from selected micromagnetic simulations to demonstrate the underlying physical theory coded in \texttt{CuPyMag} and its relevance to large-scale multiphysics applications. We first simulate the hysteresis loops and magnetic domain patterns with varying defect geometries (Fig.~\ref{Fig:domains}). These simulations align with our previous FFT-based results \cite{balakrishna2021tool,ARB2021Permalloy,renuka2022design,PhysRevMaterials.9.044407} and experimental observations \cite{hubert1998magnetic,schafer2020tomography}, which validate \texttt{CuPyMag} as a reliable platform for micromagnetic problems. Then we investigate the role of magnetoelastic interactions on coercivity and domain pattern formation in NiFe and FeGa systems, respectively (Fig.~\ref{Fig:permalloy_galfenol}). By doing so, we not only validate \texttt{CuPyMag} but also demonstrate the significant effect of magnetoelastic coupling on material response. In general, these results highlight \texttt{CuPyMag}’s ability to resolve the effects of geometry and model multiphysics interactions in large-scale micromagnetic simulations.

In our calculations, we find that simulating defects with curved boundaries (Figs.~\ref{Fig:domains},\ref{Fig:permalloy_galfenol}) requires fine meshes with at least 100k nodes for numerical convergence. These curved-boundary defects require fine discretizations to reduce artificial ``magnetic charges'' at sharp corners and to minimize errors in calculation the demagnetization field. Moreover, micromagnetic simulations generally require a mesh resolution up to the exchange length scale $l_{\mathrm{ex}}=\sqrt{\frac{2A}{\mu_0m_s^2}}$ \cite{brown1963micromagnetics}, which is a few nanometers for the NiFe alloy \cite{ARB2021Permalloy,RevModPhys.25.42}. This is because the exchange energy term only significantly penalizes magnetization gradient $\grad\vb{m}$ below this length scale. A dense mesh could also reveal a smooth magnetization rotation inside the domain walls, as shown in Figs.~\ref{Fig:domains},\ref{Fig:permalloy_galfenol}. These observations underscore the significance of \texttt{CuPyMag}'s high performance in addressing complex micromagnetic problems.

In our first example, we plot the magnetic domains for different defects at some representative points on the hysteresis loop, as shown in Fig.~\ref{Fig:domains}. The material parameters are set to be the same as that of $\mathrm{Ni_{70}Fe_{30}}$, as reported in \cite{ARB2021Permalloy,RevModPhys.25.42}. We note that for the three different defect shapes, namely a cuboid, a sphere, and an ellipsoid, the system undergoes similar needle, stripe, and needle domain transitions under an external field. This domain pattern evolution is physically expected as it minimizes the micromagnetics energy in Eq.~\ref{MicromagneticsEnergy}, in particular the anisotropy energy ($\kappa_1$ term) and the magnetostatic energy ($\vb{H}_{\mathrm{d}}$ term). They are similar to the results reported in \cite{balakrishna2021tool,ARB2021Permalloy,renuka2022design,PhysRevMaterials.9.044407} using FDM with FFT, and they reproduce the same observations in experiments that the needle domains exist for various defect shapes \cite{hubert1998magnetic,schafer2020tomography}. Together, these agreements show that \texttt{CuPyMag} captures the appropriate physics and produces consistent results in micromagnetic systems.

In Fig.~\ref{Fig:permalloy_galfenol}, we calculate the coercivity 
for the Ni-Fe alloy with different Ni compositions for all three defects. The material parameters are the same as reported in \cite{ARB2021Permalloy}. The external field and initial magnetization are imposed on the easy axis, namely $\langle 100\rangle$ crystallographic direction for $\kappa_1>0$ and $\langle 111\rangle$ for $\kappa_1<0$. The numerical implementation details of rotated coordinate systems are discussed in Appendix D. As a result, we see similar coercivity trends as previously reported \cite{ARB2021Permalloy,balakrishna2021tool}. In addition, we found that the coercivity is dependent on the defect geometry, highlighting the importance of \texttt{CuPyMag}'s FEM framework. Physically, this could be related to the domain wall pinning effect \cite{10.1063/1.4876451,Sun2021,RUIZGOMEZ2022154045}. The detailed discussion, however, is beyond the scope of this work.

In particular, we found that the magnetoelastic coupling, though computationally expensive, plays an essential role in materials research. As shown in Fig.~\ref{Fig:permalloy_galfenol}(a), without considering magnetoelastic coupling, the Ni$_{75}$Fe$_{25}$ with anisotropy constant $\kappa_1=0$ is consistently shown to have the minimum coercivity regardless of defect shape. In Fig.~\ref{Fig:permalloy_galfenol}(b), take the ellipsoid defect as an example, it is shown that the coercivity is significantly reduced for $x=0.6-0.7$ in $\mathrm{Ni}_{x}\mathrm{Fe}_{1-x}$ alloy when the magnetoelastic coupling is included. In addition, we see another minimum of coercivity at $x=0.785$ (the permalloy composition) with an uniaxial external stress $\sigma_{\mathrm{ext}}=\sigma_{11}\hat{\vb{e}}_1\otimes\hat{\vb{e}}_1 $, where $\sigma_{11} = 6$ MPa. Therefore, a suitable design of magnetoelastic response, particularly through external stress, could significantly alter the coercivity of the magnets \cite{PhysRevMaterials.9.044407}.

In addition, the external stress could also modify the magnetic domain patterns, as shown in Fig.~\ref{Fig:permalloy_galfenol}(c-d). With a compressive uniaxial stress $\sigma_{11}=-8$ MPa, the magnetic moments of the Fe$_{87}$Ga$_{13}$ alloy (galfenol) \cite{10.1063/1.3618684} are forced to rotate from $\hat{\vb{e}}_1$ to $\hat{\vb{e}}_2$ direction due to the energy penalization imposed by the $-\sigma_{\mathrm{ext}}:E$ term in Eq.~\ref{MicromagneticsEnergy}. The magnetic field behaves almost like a viscous fluid flowing in the $\hat{\vb{e}}_2$ direction, where it splits into two main “streams” that bend smoothly around the non-magnetic sphere, and rejoins on the other side. These stress-induced rotations produce domain structures that do not appear without considering magnetoelastic coupling.

\section{Discussion}
{\color{black}
\subsection{Significances and Highlights}
To emphasize \texttt{CuPyMag}’s key contributions, we summarize this subsection into four points:
\begin{itemize}
    \item \textbf{Physics modeling.}
        \texttt{CuPyMag} is developed using the micromagnetics theory with magnetoelastic interactions. Additionally, \texttt{CuPyMag} uses the ellipsoid theorem in computing long-range field interactions (e.g., demagnetization field) and accounts for localized non-ferromagnetic defects. These features enable \texttt{CuPyMag} to be an effective mesoscale tool to explore the effect of defect shapes on magnetic domain wall dynamics and hysteresis. These advancements align closely with the growing trend toward multiphysics coupling in modern micromagnetic simulations \cite{10.1063/1.5093730}. Such physics modeling helped us identify the minimum coercivity at the permalloy composition \cite{ARB2021Permalloy,RevModPhys.25.42} and the stress-induced domain change in Galfenol \cite{Guevara2022}.
    \item \textbf{Computational efficiency.}
        \texttt{CuPyMag} uses Numba to accelerate the initial system assembly, while subsequent coupled LLG loop calculations are executed efficiently in a tensorized, BLAS-accelerated GPU workflow enabled by CuPy. As demonstrated in our benchmark tests, this design sustains linear-sublinear runtime growth with problem size up to the upper bound imposed by the GPU memory. A hysteresis simulation of a coupled system with 3 million nodes was completed within 3 hours on a single H200 GPU, achieving a 2-3$\times$ double-precision speedup compared to A100. 
        By minimizing CPU-GPU data transfers and exploiting modern GPU architectures, \texttt{CuPyMag} can handle large-scale systems with complex boundary geometries on an unstructured FEM mesh.
    \item \textbf{User accessibility.}
        \texttt{CuPyMag} uses widely adopted Python libraries such as Numba and CuPy, which are compatible with Conda environments and installable via \texttt{conda} or \texttt{pip}. The entire project can be installed with a single \texttt{pip install -e .} command, ensuring rapid environment setup across different operating systems, such as Linux and Windows. Simulations are configured through a single, intuitive YAML file, lowering the barrier for adoption in materials research. 
    \item \textbf{Mathematical transparency and adaptability.}
        \texttt{CuPyMag} avoids unnecessary abstraction layers and directly mirrors FEM mathematical theory in its modular architecture. Governing equations are expressed in the standard linear algebra form $A\vb{x=b}$, and key operations such as matrix assembly, boundary condition enforcement, and volume averaging are implemented as concise, readable functions. This one-to-one mapping between mathematical concepts and code modules facilitates debugging, verification against analytical solutions, and straightforward extension to new interactions or physics systems. Therefore, in addition to being a micromagnetic simulation program, \texttt{CuPyMag} also serves as a transparent and efficient tool for algorithm developers. For readers who wish to adapt the code as a more general FEM solver, illustrative example scripts are included in the repository’s examples directory.
\end{itemize}
\subsection{Limitations and Potential Future Developments}
Next, we identify \texttt{CuPyMag}'s limitations and potential directions for its future development:

\begin{itemize}
  \item \textbf{Preconditioned solvers.}  
    At present, \texttt{CuPyMag} employs an unpreconditioned GG solver, which is GPU-efficient due to high memory-bandwidth utilization (highly parallel). However, the mechanical-equilibrium systems in Fig.~\ref{Fig:GSPM} require hundreds of CG iterations to converge in the given tolerance, indicating a bottleneck for the runtime. Incorporating GPU-efficient preconditioners via external libraries such as \texttt{pyamgx} \cite{pyamgx} or \texttt{petsc4py} \cite{DALCIN20111124,osti_2565610} without significantly increasing GPU memory demand (as \texttt{CuPyMag} is already memory-bound) would significantly reduce iteration count and accelerate the solver.

  \item \textbf{Higher‑order and curved elements.}  
    Currently, only linear hexahedral and tetrahedral Lagrange elements are supported in \texttt{CuPyMag}. Extending this to higher‑order or \(C^1\)‑continuous elements, and curved‑boundary elements \cite{doi:10.1137/0710022}, would enhance the geometric flexibility and simulation fidelity of our open-source framework.

  \item \textbf{Additional boundary conditions.}  
    The present framework assumes periodic boundary conditions, which narrows the class of problems that can be modeled. The framework can be further generalized by implementing Neumann boundary conditions and adding surface integrals in the weak form to the magnetostatic equilibrium equation (Eq.~\ref{weak_demag}). This extension would broaden the range of physical problems that can be addressed by \texttt{CuPyMag}.

  \item \textbf{Assembly‑free approaches.}  
    \texttt{CuPyMag}’s current FEM implementation assembles and stores sparse matrices explicitly, leading to large GPU memory demand and makes mesh deformation at each timestep prohibitively expensive to consider. By developing an assembly-free FEM framework on GPU \cite{10.1115/1.4028591,Kiran2020,MARTINEZFRUTOS20159}, we would increase memory efficiency and enable simulations with evolving meshes.

  \item \textbf{Multi‑GPU scalability.}  
    At present, \texttt{CuPyMag}’s simulations are limited to a single GPU, which constrains problem size and performance. Extending the framework with multi-GPU support, using CUDA‑aware MPI \cite{haidar2012mvapich2-gdr} or NVIDIA NCCL \cite{nccl} would dramatically increase simulation capacity. This would allow users to apply \texttt{CuPyMag} to investigate large-scale material problems, including nanocrystalline materials \cite{HERZER1992258,HERZER2013718}, magnetostrictive polycrystals \cite{GAO2022177}, and crack‑driven domain‑wall dynamics \cite{RojasDiaz2009}.
\end{itemize}
}

\section{Conclusion}
In this work, we presented \texttt{CuPyMag}, an open‑source, Python‑based finite‑element micromagnetic simulation framework that incorporates magnetoelastic coupling, the ellipsoid theorem, and arbitrary-shaped defects. By leveraging Numba for the assembly of the sparse system matrices on the CPU and subsequent tensorized, BLAS-accelerated operations for a GPU-resident workflow that minimizes CPU-GPU communications, \texttt{CuPyMag} achieves a 3-hour hysteresis loop simulation of a system with 3 million nodes on a single H200 GPU. Performance benchmarks demonstrate that our Numba-accelerated assembly scales linearly with problem size, while the subsequent GPU routines scale linearly to sublinearly with problem size. Furthermore, the GSPM maintains numerical stability with time steps up to 11 ps, requiring only 1-3 CG iterations per solve. Simulations on cuboid, spherical, and ellipsoidal defects reproduce similar needle-to-stripe-to-needle transitions reported previously, and capture stress-induced coercivity and magnetic domain changes. Using widely adopted Python libraries, \texttt{CuPyMag} combines high-level programmability and readability with GPU speed, ensuring cross-platform compatibility and is readily extensible for new physics through its modular, theory-aligned architecture.

\section*{CRediT authorship contribution statement}
\textbf{Hongyi Guan:} Conceptualization, Methodology, Software, Formal analysis, Investigation, Data curation, Visualization, Project administration, Writing – original draft, Writing – review \& editing. \textbf{Ananya Renuka Balakrishna:} Conceptualization, Supervision, Project administration, Funding acquisition, Writing – review \& editing.

\section*{Declaration of competing interest}
The authors declare that they have no known competing financial interests or personal relationships that could have appeared to influence the work reported in this paper.

\section*{Acknowledgement}
The authors acknowledge research funding from the U.S. Department of Energy (DOE), Office of Basic Energy Sciences, Division of Materials Sciences and Engineering under Award DE-SC0024227 (Code development, Computational studies; HG, ARB). The authors acknowledge the use of computational resources provided by both the Center for Scientific Computing at University of California, Santa Barbara (MRSEC; NSF DMR 2308708) and the Delta system at the National Center for Supercomputing Applications through allocation MAT250025 from the Advanced Cyberinfrastructure Coordination Ecosystem: Services \& Support (ACCESS) program, which is supported by National Science Foundation grants \#2138259, \#2138286, \#2138307, \#2137603, and \#2138296.

\newpage

\pretocmd{\appendix}{%
  \counterwithin{equation}{section}%
  \counterwithin{table}{section}%
  \counterwithin{figure}{section}%
  \counterwithin{algorithm}{section}%
  \renewcommand{\theequation}{\thesection.\arabic{equation}}%
  \renewcommand{\thetable}{\thesection.\arabic{table}}%
  \renewcommand{\thefigure}{\thesection.\arabic{figure}}%
  \renewcommand{\thealgorithm}{\thesection.\arabic{algorithm}}%
  \renewcommand{\tablename}{Table}%
  \renewcommand{\figurename}{Figure}%
}{}{}

\appendix

\section{Material Constants and Notations}
\begin{table}[!h]
\centering
\begin{tabular}{p{2cm} p{12cm}}
\hline 
Notation & Description \tabularnewline
\hline

$\mu_0$ & Vacuum permeability constant \tabularnewline

$m_s$ & Saturation magnetization in the ferromagnetic phase \tabularnewline

$\vb{m}$ & Nondimensionalized magnetization, $\vb{m} = \frac{\vb{M}}{m_s}$ \tabularnewline

$A$ & Gradient energy (or exchange energy) coefficient. We set $A$ as a scalar in \texttt{CuPyMag}, but generally it is a matrix. \tabularnewline

$\kappa_1$ & Magnetocrystalline anisotropy constant associated with rotation the magnetization away from its easy axes  \tabularnewline

$\lambda_{100}$, $\lambda_{111}$  & Magnetostriction constants represent the relative change in length of a magnetic material when exposed to an external field. The subscripts denote the strains along the $\langle 100\rangle$ and the $\langle 111\rangle$ crystallographic directions.\tabularnewline

$\vb{u}$ & Strain displacement \tabularnewline

$E$ & Strain tensor \tabularnewline

$E_0(\vb{m})$ & Spontaneous (or preferred) strain tensor. For a cubic crystal it is calculated as:
\[  
E_0(\vb{m}) = \dfrac{3}{2}\mqty(\lambda_{100}\mathrm{(m_1^2-1/3)} & \mathrm{\lambda_{111}m_1m_2} & \mathrm{\lambda_{111}m_1m_3} \\ \mathrm{\lambda_{111}m_1m_2} & \lambda_{100}\mathrm{(m_2^2-1/3)} & \mathrm{\lambda_{111}m_2m_3} \\ \mathrm{\lambda_{111}m_1m_3} & \mathrm{\lambda_{111}m_2m_3} & \lambda_{100}\mathrm{(m_3^2-1/3)}) \]
\vspace{-0.3cm}
\tabularnewline

$\sigma_{\mathrm{ext}}$ & External stress \tabularnewline

$\mathbb{C}$ & Elastic stiffness constant \tabularnewline

$U$ & Magnetostatic potential: 
\[
U(\vb{x}) = -\frac{1}{4\pi}\int \grad \dfrac{1}{|\vb{x-y}|}\cdot \vb{M(y)}\dd \vb{y}\]
\vspace{-5mm}
\tabularnewline 

$\vb{H}_{\mathrm{d}}$ & Demagnetization field, $\vb{H}_{\mathrm{d}} = -\grad U$ \tabularnewline

$\vb{H}_{\mathrm{ext}}$ & External field \tabularnewline

\hline
\end{tabular}
\caption{Notations and description of symbols used in the text} \label{Symbols}
\end{table}

\section{The Gauss-Seidel Projection Method}
We minimize the free energy functional $\Psi[\vb{m}]$ (Eq.~\ref{MicromagneticsEnergy}) via the Landau-Lifshitz-Gilbert (LLG) equation \cite{LandauLifshitz1935,1353448}:
\begin{equation}
    \pdv{\vb{m}}{\tau} = -\vb{m}\times \vb{H}_{\mathrm{eff}}-\alpha\vb{m}\times(\vb{m}\times\vb{H}_{\mathrm{eff}}),
\end{equation}
where $\tau = \gamma m_s t$ is the dimensionless timestep, and $\gamma$ is the gyromagnetic ratio. $\alpha$ is the damping constant. The effective field is the variational derivative of the free energy functional:
\begin{equation}
    \vb{H}_{\mathrm{eff}} = -\dfrac{1}{\mu_0m_s^2}\fdv{\Psi}{\vb{m}} = A^{*}\grad^2\vb{m} + \vb{h(m)}.
\end{equation}
The first term is related to the gradient term of $\vb{m}$, and $A^{*} = 2A/\mu_0m_s^2l_d^2$ is the reduced exchange energy coefficient, with $l_d$ being the length scale of the model. The second term is the variational derivative for the remaining term. The Gauss-Seidel projection method updates the $\vb{m}$ is the following way:
\begin{algorithm}[ht]
  \caption*{\textbf{Algorithm B.1:} The Gauss–Seidel Projection Method}
  \label{alg:gs}
  \begin{algorithmic}[1]
    \Require Initial magnetization $\vb{m}^0(\vb{x})$, time step $\Delta\tau$, damping parameter $\alpha$, constant $A^*$
    \Ensure Sequence $\{\vb{m}^n\}$ converging to equilibrium
    \For{$n=0,1,2,\ldots$ until convergence}
      \State \textbf{Step 1:} Compute intermediate fields
      \[
        \vb{g}^n(\vb{x}) = (1 - A^* \Delta\tau\nabla^2)^{-1}\bigl[\vb{m}^n(\vb{x})
          + \Delta\tau\vb{h}[\vb{m}^n]\bigr],
      \]
      \[
        \vb{g}^*(\vb{x}) = (1 - A^* \Delta\tau\nabla^2)^{-1}\bigl[\vb{m}^*(\vb{x})
          + \Delta\tau \vb{h}[\vb{m}^n]\bigr].
      \]
      \State \textbf{Step 2:} Update magnetization $\vb{m}^*$ by
      \[
        \vb{m}^* =
        \begin{bmatrix}
          m^n_1 + (g^n_2m^n_3 - g^n_3m^n_2)\\[4pt]
          m^n_2 + (g^n_3m^n_1 - g^n_1m^n_3)\\[4pt]
          m^n_3 + (g^*_1m^*_2 - g^*_2m^*_1)
        \end{bmatrix}.
      \]
      \State \textbf{Step 3:} Compute damped field
      \[
        \vb{m}^{**}
        = (1 - A^* \alpha \Delta\tau\nabla^2)^{-1}
          \bigl[\vb{m}^* + \alpha \Delta\tau \vb{h}[\vb{m}^*]\bigr].
      \]
      \State \textbf{Step 4:} Enforce unit‐length constraint
      \[
        \vb{m}^{n+1} \gets \frac{\vb{m}^{**}}{\lVert \vb{m}^{**}\rVert}.
      \]
    \EndFor
  \end{algorithmic}
\end{algorithm}

We can see that all of the equations we have to solve are of the same mathematical structure as Eq.~\ref{GS}.

\section{Sample Illustration of Our FEM Mesh}
\begin{figure}[H]
    \centering
    \includegraphics[width=\linewidth]{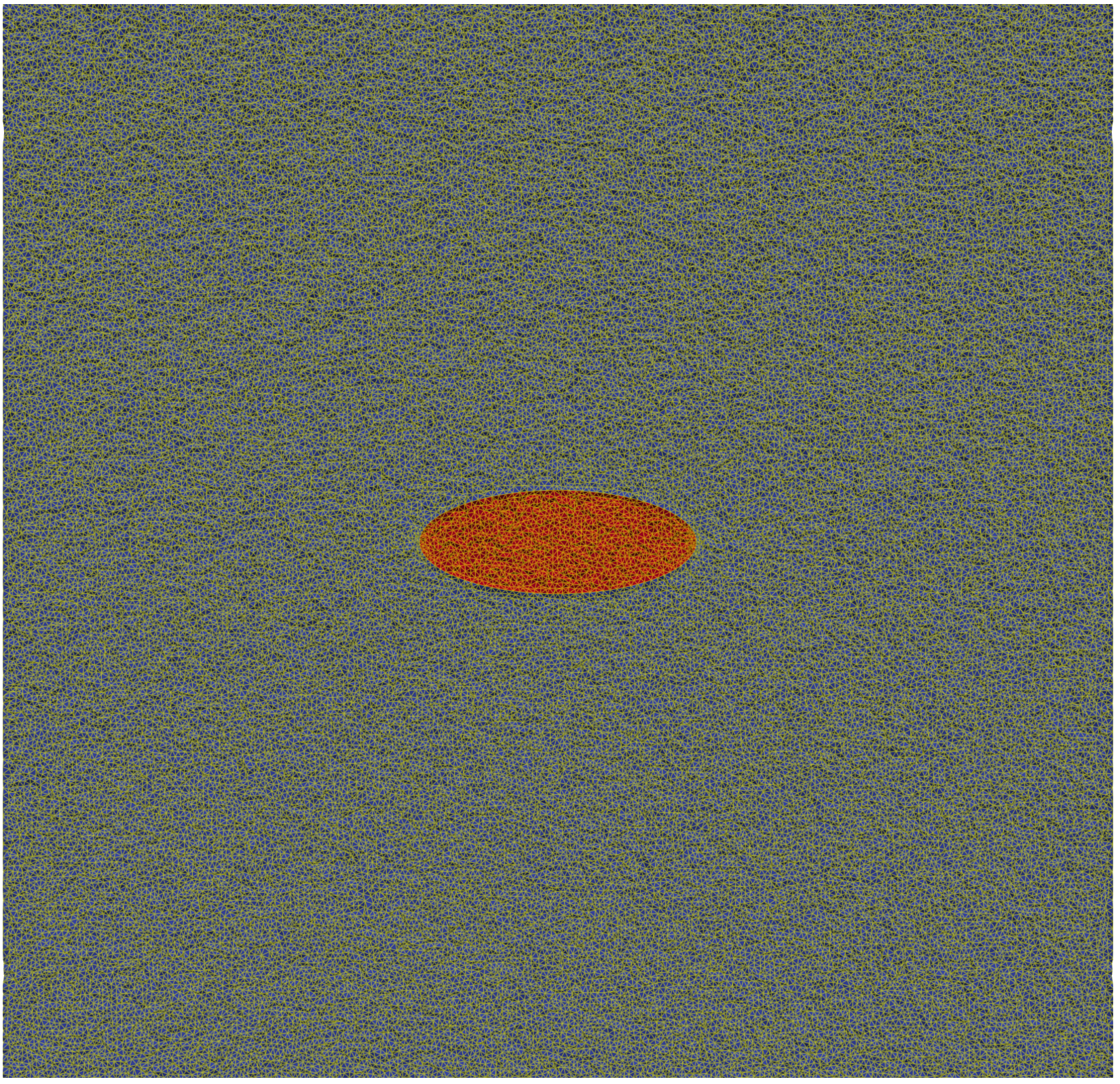}
    \caption{An illustration of the most dense mesh used in our benchmark (Mesh 6 in Figure~\ref{Fig:TotalBenchmark}(h)).}
    \label{Fig:prism_mesh}
\end{figure}
\begin{figure}[H]
    \centering
    \includegraphics[width=\linewidth]{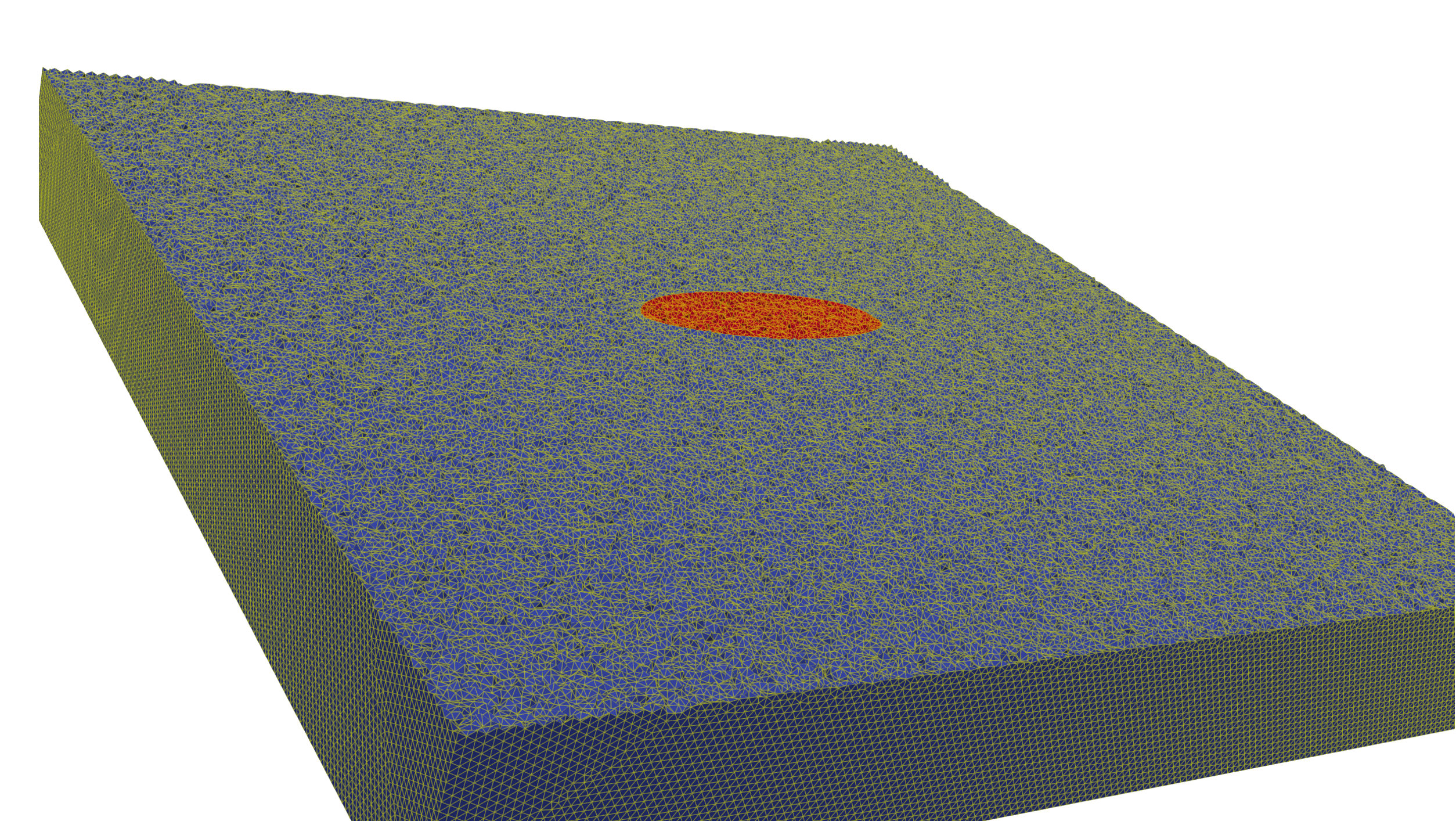}
    \caption{An illustration of the most dense mesh used in our benchmark (Mesh 6 in Figure~\ref{Fig:TotalBenchmark}(h)), from a different angle.}
    \label{Fig:sphere_mesh}
\end{figure}

\section{Simulations under Rotated Coordinate Systems}
The rotated coordinate has 3 orthogonal principal axes along $\langle 111\rangle$, $\langle \bar{1}10 \rangle$, and $\langle 11\bar{2} \rangle$, thus follows the rotation matrix:
\begin{equation}
    R=\mqty(1/\sqrt{3} & 1/\sqrt{3} & 1/\sqrt{3} \\ -1/\sqrt{2} & 1/\sqrt{2} & 0 \\ -1/\sqrt{6} & -1/\sqrt{6} & 2/\sqrt{6}).
\end{equation}
There, for any physical quantity $\vb{v}\in\mathbb{R}^3$, and $A\in \mathbb{R}^{3\times3}$, we have
\begin{equation}
    \vb{v}_{111} = R\vb{v}_{100}, ~ A_{111} = RA_{100}R^{\operatorname{T}}
\end{equation}
For Voigot notation, we have $E\in\mathbb{R}^6$ (a strain), and $\mathbb{C}\in\mathbb{R}^{6\times6}$ (an elastic stiffness tensor), they follow a rotation matrix
\begin{equation}
    M = \mqty(R_{xx}^{2} & R_{xy}^{2} & R_{xz}^{2} &
    2 R_{xy} R_{xz} & 2 R_{xx} R_{xz} & 2 R_{xx} R_{xy}\\
    R_{yx}^{2} & R_{yy}^{2} & R_{yz}^{2} & 2 R_{yy} R_{yz} &
    2 R_{yx} R_{yz} & 2 R_{yx} R_{yy}\\
    R_{zx}^{2} & R_{zy}^{2} & R_{zz}^{2} & 2 R_{zy} R_{zz} &
    2 R_{zx} R_{zz} & 2 R_{zx} R_{zy}\\
    R_{yx} R_{zx} & R_{yy} R_{zy} & R_{yz} R_{zz} & R_{yy}
    R_{zz} + R_{yz} R_{zy} & R_{yx} R_{zz} + R_{yz} R_{zx}
    & R_{yx} R_{zy} + R_{yy} R_{zx}\\
    R_{xx} R_{zx} & R_{xy} R_{zy} & R_{xz} R_{zz} & R_{xy}
    R_{zz} + R_{xz} R_{zy} & R_{xx} R_{zz} + R_{xz} R_{zx}
    & R_{xx} R_{zy} + R_{xy} R_{zx}\\
    R_{xx} R_{yx} & R_{xy} R_{yy} & R_{xz} R_{yz} & R_{xy}
    R_{yz} + R_{xz} R_{yy} & R_{xx} R_{yz} + R_{xz} R_{yx}
    & R_{xx} R_{yy} + R_{xy} R_{yx}).
\end{equation}
Here we have a similar transform:
\begin{equation}
    E_{111} = ME_{100}, ~ \mathbb{C}_{111} = M\mathbb{C}_{100}M^{\operatorname{T}}.
\end{equation}
Therefore, the following routine is used for a rotated coordinate system:

\begin{algorithm}[ht]
  \caption*{\textbf{Algorithm D.1:} The Rotated Coordinate System}
  \label{alg:rt}
  \begin{algorithmic}[1]
    \Require Initial magnetization $\vb{m}^0(\vb{x})$, rotation matrices $R$, $M$.
    \State Compute $\mathbb{C}_{111} = M\mathbb{C}_{100}M^{\operatorname{T}}$ and set up the FEM system.
    \For{Each LLG step $n$}
        \State \textbf{Step 1:} Compute \[E_{0,111}(\vb{m}_{111}) = ME_{0,100}(\vb{m}_{100}), \] where $\vb{m}_{100}=R^{-1}\vb{m}_{111}$, and the form is $E_{0,100}(\vb{m}_{100})$ is shown in Table~\ref{Symbols}.
        \State \textbf{Step 2:} Solve the mechanical equilibrium \[\div \mathbb{C}_{111}[E_{111}-E_0(\vb{m}^n_{111})]=0,\] and transfer to $\langle 100 \rangle$ coordinate system: $E_{100} = M^{-1}E_{111}$.
        \State \textbf{Step 4:} Compute  
        \[ \vb{h}_{100}(\vb{m}_{100}, E_{100}) = \pdv{\Psi}{\vb{m}_{100}}, \]
        and obtain the derivatives at $\langle 111\rangle$ coordinate systems by
        \[
            \vb{h}_{111} = \pdv{\Psi}{\vb{m}_{111}} = \pdv{\Psi}{\vb{m}_{100}}\cdot\pdv{\vb{m}_{100}}{\vb{m}_{111}} = R\vb{h}_{100}.
        \]
    \EndFor
  \end{algorithmic}
\end{algorithm}
Other steps are the same as in the original coordinate system.

\bibliographystyle{elsarticle-num}
\bibliography{references}

\end{document}